\documentclass[preprint2]{aastex}

\newcommand{\DDt}[1]{ \frac{\mathrm{D} #1}{\mathrm{D} t} }
\newcommand{\ddt}[1]{ \frac{\mathrm{d} #1}{\mathrm{d} t} }
\newcommand{\ddthat}[1]{ \frac{\mathrm{d} #1}{\mathrm{d} \hat{t}} }

\newcommand{\partials}[2]{ \frac{\partial #1}{\partial #2} }

\usepackage{natbib}
\usepackage{multirow}
\usepackage{rotating}
\usepackage{amsmath}

\usepackage{array}
\newcolumntype{T}{c}
\usepackage{float}

\begin{document}

\title{Deciphering thermal phase curves of dry, tidally locked terrestrial planets}
\author{Daniel D.~B.~Koll and Dorian S.~Abbot}
\affil{Department of the Geophysical Sciences, University of Chicago,
    Chicago, IL 60637}
\email{dkoll@uchicago.edu}

\begin{abstract}
  Next-generation space telescopes will allow us to characterize
  terrestrial exoplanets. To do so effectively it will be crucial to
  make use of all available data. We investigate which atmospheric
  properties can, and cannot, be inferred from the broadband
    thermal phase curve of a dry and tidally locked terrestrial
    planet. First, we use dimensional analysis to show that phase
  curves are controlled by six nondimensional parameters. Second, we
  use an idealized general circulation model (GCM) to explore the
  relative sensitivity of phase curves to these parameters.  We find
  that the feature of phase curves most sensitive to atmospheric
  parameters is the peak-to-trough amplitude. Moreover, except for hot
  and rapidly rotating planets, the phase amplitude is primarily
  sensitive to only two nondimensional parameters: 1) the ratio of
  dynamical to radiative timescales, and 2) the longwave optical depth
  at the surface.  As an application of this technique, we show how
  phase curve measurements can be combined with transit or emission
  spectroscopy to yield a new constraint for the surface pressure and
  atmospheric mass of terrestrial planets.  We estimate that a single
  broadband phase curve, measured over half an orbit with the
  \textit{James Webb Space Telescope}, could meaningfully constrain
  the atmospheric mass of a nearby super-Earth.  Such constraints will
  be important for studying the atmospheric evolution of terrestrial
  exoplanets as well as characterizing the surface conditions on
  potentially habitable planets.
\end{abstract}


\section{Introduction}
\label{sec:introduction}

%% IS THERE A BETTER WAY THAN SLOPPINESS? xxx
\sloppy

Data from the Kepler telescope indicate that $\sim50$-$100\%$ of nearby
cool stars host a rocky planet
\citep{dressing2013,morton2014}. If we were able to characterize even
a fraction of these planets we could vastly expand our understanding
of processes fundamental for terrestrial planets, including planet
formation, atmospheric escape, photochemistry, and atmospheric
dynamics.  The observational best-case scenario is a transiting
planet, whose orbit we happen to view edge-on, so that the planet
periodically passes in front of and behind its star. Broadly speaking,
to characterize such a planet we would want to determine its atmospheric
composition, temperature structure, and atmospheric mass\footnote{The
  mass of an atmospheric column with unit surface area is $p_s/g$,
  where $p_s$ is the surface pressure and $g$ is the surface
  gravity.}.  The composition reflects how the planet formed, how its
atmosphere subsequently evolved (e.g., via degassing from the interior
or atmospheric escape), and the chemical state of its
atmosphere. The temperature structure indicates the dynamical regime
of the atmosphere and, if retrievable down to the surface, whether the
planet could be habitable. The atmospheric mass reflects the planet's
atmospheric evolution and also determines its habitability (by
controlling whether water can exist as a liquid).

The most mature techniques for characterizing transiting planets are
transit spectroscopy, in which starlight is measured as it filters
through a planet's atmosphere, and emission spectroscopy, in which a
planet's thermal emission is measured just before the planet is
occulted by its star. In theory, high-resolution transit and emission
spectra both contain enough information to uniquely constrain atmospheric
composition, temperature structure, atmospheric mass, and planetary mass
\citep{madhusudhan2009,benneke2012,line2012,lee2012,wit2013}.
In practice, it is difficult to comprehensively characterize even
hot Jupiters with any single spectroscopic technique due to
measurement error and observational degeneracies
\citep[e.g.,][]{burrows2013,hansen2014,griffith2014}.

It is therefore desirable to seek additional methods for
characterizing terrestrial planets that complement high-resolution
spectroscopy. One simple approach is to observe a planet's broadband
thermal phase curve, which is the net infrared flux the planet emits
as it orbits its star. Before the planet passes in front of its star
we observe flux emitted from the planet's nightside, and just before
the planet passes behind its star we observe flux from the planet's
dayside. The resulting phase curve can then be used to infer five pieces of
information: the planet's average thermal emission, the
location of hot and cold spots and the flux emitted at the hot and
cold spots \citep{cowan2008}.  This technique has already been applied
to hot Jupiters.
For example, \citet{knutson2007} were able to infer
equatorial superrotation on HD 189733b from the fact that its hot spot
is shifted eastward of the substellar point, consistent with the
theoretical prediction of \citet{showman2002}.
It will be more challenging to measure thermal phase curves of smaller and cooler
planets, but it should be possible to perform such measurements using
next-generation instruments like the \textit{James Webb Space
  Telescope} \citep[\textit{JWST};][]{deming2009b}.

Although thermal phase curve measurements of terrestrial planets will
soon be technically feasible, more work is needed to determine how
they can be fully exploited. A natural starting point is to assume
that planets accessible to near-future observations will be tidally
locked (even though planets could also be trapped in higher-order spin
resonances, see Section \ref{sec:discussion}). For a 
tidally locked planet the phase curve depends largely on the
atmospheric redistribution of energy between dayside and
nightside. Many researchers have therefore proposed using phase curves
to characterize tidally locked planets
\citep{seager2009,cowan2011,selsis2011,menou2012a,yang2013,mills2013,perez-becker2013a,yang2014,kataria2014a}. At
the same time, these results have also shown that phase curves are
sensitive to multiple atmospheric parameters, which makes them
difficult to interpret. For example, a small phase curve amplitude is
compatible with: 1) a massive atmosphere because thicker atmospheres
transport heat more effectively \citep{selsis2011}, 2) an atmosphere
containing large amounts of H$_2$, which has a higher heat capacity
than high mean-molecular-weight gases and therefore loses heat more
slowly as air is advected to the nightside \citep{menou2012a}, 3)
relatively weaker absorption of shortwave radiation, so that stellar
energy is deposited at higher pressures before being reemitted to
space \citep{burrows2010,heng2011}, and 4) a low magnetic drag in
ionized atmospheres, which allows higher wind speeds and thus more
efficient heat transport \citep{rauscher2012}.

In this paper we disentangle how different atmospheric parameters
affect phase curves and show how the phase curve amplitude can be used
to constrain atmospheric mass.  We focus on the phase curve amplitude
because we find that, for many terrestrial planets, hot/cold spot
offsets will be small (see Section \ref{sec:results}).  We focus on
atmospheric mass because it will be difficult to infer from either
transit or emission spectroscopy, as can be seen from the following argument. Following
\citet{lecavelierdesetangs2008}, the maximum pressure that can be
probed in transit is $p_{max} = 0.56 \times g/\kappa_{min} \times
\sqrt{H/(2 \pi a)}$, where $g$ is the acceleration of gravity,
$\kappa_{min}$ is the opacity per unit mass in the most transparent
part of the spectrum, $H$ is the scale height, and $a$ is the
planetary radius\footnote{Compared to Section 4.1 in
  \citet{lecavelierdesetangs2008}, we additionally define $\kappa
  \equiv \sigma_0 / \mu$, $R\equiv k/\mu$ and $H\equiv R T/g$.}.  If
we assume that Rayleigh scattering dominates the transit spectrum up
to $0.75~\mu$m and that this is the most transparent part of the
spectrum, then, for an N$_2$ atmosphere, $\kappa_{min} \sim 2.59
\times 10^{-6}$ m$^2$ kg$^{-1}$ \citep[Table
5.2,][]{pierrehumbert2011b}. For an Earth analog with $a=a_\Earth$,
$g=10$ m s$^{-2}$ and $H=8$ km, we find that $p_{max} = 0.3$ bar.  In
reality it would be even harder to probe an atmosphere this deeply in
transit due to clouds and hazes \citep{fortney2005} or atmospheric
refraction \citep{betremieux2013,misra2014b}.
Emission spectroscopy generally probes deeper into an atmosphere than
transit spectroscopy. Atmospheric mass can then be constrained using
the fact that pressure-broadening widens molecular absorption features at
higher pressures. Nevertheless, pressure and molecular abundances are
largely degenerate in their effect on emission spectra, which
complicates the interpretation of emission spectra. For example,
\citet{vonparis2013} estimate that emission spectroscopy of a cool
Earth analog with a $1$ bar atmosphere could place an upper bound on the surface pressure of
about $5.6$ bar.
Obtaining the upper bound would require a low-resolution
spectrum ($\lambda/\Delta\lambda = 20$) with a signal-to-noise ratio
($SNR$) of 10 (their Table 3).
To estimate how much observation time
this would require on \textit{JWST}, we use the $SNR$ maps in
\citet{belu2011} as a guideline. We estimate that, for a cool M-dwarf
at $5$ pc, $SNR=10$ at this
spectral resolution would require $\sim14$ days of continuous
monitoring\footnote{We assume $SNR=10$ for the
  detection of $CO_2$ at $15~\mu$m \citep[Fig.~16b in][]{belu2011} is representative
  for the entire thermal range, which is optimistic (compare to their
  Fig.~15b). To allow comparison with Section \ref{sec:results}, we
  additionally assume a 0.2 solar mass host star and rescale the
  required observation time for a target at $10$ pc, which is $\sim
  3\%$ of the \textit{JWST} main mission or $\sim1.8$ months,
  to a target at $5$ pc.}. This amounts to observing roughly every
eclipse the planet makes during \textit{JWST}'s five-year mission lifetime.
The pressure-dependent formation of O$_2$ dimers offers another method
for measuring atmospheric mass \citep{misra2014a}. However, being able
to detect the dimer spectral signature requires an atmosphere with O$_2$
concentrations similar to Earth's.
The observation time necessary would again amount to $\sim 12$ days of continuous
observation, or almost all available transits over \textit{JWST}'s
mission lifetime. Such long-term and detailed observations could be feasible for high-priority
observation targets, but even in those cases it would be desirable to
have an independent and less time-consuming way of estimating the
atmospheric mass.

In the following sections we first analyze the dynamical and radiative
equations relevant for terrestrial planet atmospheres. We specifically
consider atmospheres that are ``dry'' (i.e., condensation is negligible) and tidally
locked. For such atmospheres we identify six nondimensional parameters that \textit{could}
influence phase curves (Section \ref{sec:methods}).
Next, we use an idealized GCM to numerically test which of the six
parameters phase curves actually are sensitive to (Section
\ref{sec:results}). Except for hot and rapidly rotating planets with
optically thick atmospheres, we do not find significant hot/cold spot
offsets. We therefore focus on how the phase curve amplitude can be
used to constrain an atmosphere's properties. We find that, except for
hot and rapidly rotating planets, the phase curve amplitude is mainly
sensitive to two nondimensional parameters.  We then show how the
phase amplitude can be combined with information from transit or
emission spectroscopy to constrain the atmospheric mass of a
terrestrial planet (Section \ref{sec:observations}). We estimate that
one measurement of a nearby super-Earth's phase amplitude with
\textit{JWST}, taken over half the planet's orbit, could constrain the
atmospheric mass to within a factor of two.

\section{Methods}
\label{sec:methods}
We adopt a basic, yet comprehensive, model for the phase curve of a
terrestrial planet. We assume the planet is tidally locked and in a
circular orbit. This regime is particularly relevant for planets
orbiting smaller main-sequence stars, that is, K- and M-dwarfs. At
minimum, the phase curve of such a planet is set by atmospheric fluid
dynamics, radiative transfer, and surface-atmosphere exchange of
energy and momentum.  As is standard for planetary atmospheres, we
model the atmospheric fluid dynamics using the primitive equations
\citep{vallis2006}. The primitive equations assume hydrostatic
equilibrium and that horizontal length scales are much larger than
vertical ones, both of which tend to be excellent approximations for
large-scale motions. We focus on atmospheres cooler than $1000$
K, for which magnetic effects should be negligible \citep{menou2012b}.
We use bulk aerodynamic formulae for the surface exchanges of energy and
momentum.
We model the radiative transfer as two-band (shortwave and longwave) gray radiation.
  We neglect scattering and assume that longwave and
  shortwave opacities increase linearly with pressure. The linear
  dependency approximates the effects of pressure
  broadening and continuum absorption in a well-mixed atmosphere
  \citep{robinson2014b}.

We assume that the thermodynamics are dry. This is a natural
starting point for a theoretical investigation, but our results should
apply to a wide range of actual atmospheres. First, we expect that
many terrestrial planets will be dry because post-formation delivery
of volatiles via planetesimals and comets is a stochastic process
\citep{morbidelli2000}. In addition, for planets hotter than Earth,
volatiles can be lost via atmospheric escape \citep[the so-called
moist greenhouse;][]{kasting1988}. On tidally locked planets,
volatiles can also become cold-trapped on the nightside
\citep{leconte2013, menou2013}.  Moreover, the dry regime is a useful
approximation even for atmospheres like Earth's with moderate amounts
of a condensing substance \citep{schneider2006}.  We therefore expect
that insight gained in the dry regime will carry over to the moist
case. For example, if the atmospheric dynamics were insensitive to one
parameter in the dry regime (e.g., surface friction), this suggests
that moist atmospheres could be similarly insensitive.  Finally,
observations will be able to control for cases in which our analysis
no longer applies. For example, condensation and cloud formation would
lead to anomalously high bond albedos and could also reverse the
expected day-night phase curve pattern \citep{yang2013}.

The equations of our assumed model, shown in Appendix
\ref{sec:appendix1}, contain twelve dimensional parameters.
The parameters are: stellar constant, $L_*$, planetary albedo, $\alpha$,
rotation rate, $\Omega$, planetary radius, $a$, surface gravity, $g$,
specific heat capacity, $c_p$, specific gas constant, $R$, shortwave and longwave
opacities at some reference pressure, $\kappa_{SW}$ and $\kappa_{LW}$,
the Stefan-Boltzmann constant, $\sigma_{SB}$, surface pressure, $p_s$, and
surface drag coefficient, $C_D$.
For a single gas species the specific gas constant is $R \equiv k_B /(m_p M)$, where $k_B$
is the Boltzmann constant, $m_p$ is the mass of a proton, and $M$ is
the molecular weight of a gas molecule. For multiple species in a
well-mixed dry atmosphere one can similarly assign bulk
values of $R$ and $c_p$ \citep{caballero2014}.
The opacities $\kappa_{SW}$ and
$\kappa_{LW}$ are defined at a reference pressure, $p_0$. The choice
of $p_0$ is arbitrary and one could set it equal to $p_s$, so it does
not provide an additional dimensional parameter. To better
compare our choices of $\kappa_{SW}$ and $\kappa_{LW}$ with previous
work, we keep $p_0$ and $p_s$ distinct. We also note that $L_*$,
$\alpha$, and $\sigma_{SB}$ are not independent. The product $L_*(1-\alpha)$
only appears in the stellar forcing term of the radiative equations, and
$\sigma_{SB}$ only appears in the radiative equations (Appendix
\ref{sec:appendix1}). We account for the degeneracy between $L_*$,
$\alpha$, and $\sigma_{SB}$ by defining a characteristic temperature,
$T_{eq} = [L_*(1-\alpha)/(4\sigma_{SB})]^{1/4}$, which is the equilibrium
emission temperature of a spatially homogeneous planet.  This reduces
the number of dimensional parameters to ten.

Following \citet{frierson2005}, we use the Buckingham-Pi
theorem to express ten dimensional parameters measured
in four different units (mass, length, time, and temperature) as only
six nondimensional parameters \citep{buckingham1914}. There is no unique choice for these nondimensional parameters; we form them using characteristic
scales that we consider most appropriate for relatively
slowly rotating tidally locked atmospheres.
Our choice of scales nevertheless leads to
  nondimensional parameters that are well-known in the
  literature.
As a characteristic velocity scale we choose the speed of gravity
waves, $c \sim N H$, where $N$ is the Brunt-V\"ais\"al\"a\ frequency
and $H \equiv R T_{eq}/g$ is the scale height. Adjustment via gravity
waves is key in setting the day-night temperature gradients, and hence
phase curves, of relatively slowly rotating planets
\citep{perez-becker2013a,showman2013b}.  The Brunt-V\"ais\"al\"a\ frequency is given by
$N^2=g/T(g/c_p+\mathrm{d}T/\mathrm{d}z)$. The lapse rate,
$\mathrm{d}T/\mathrm{d}z$, is a priori unknown for any atmosphere. To
place an upper bound on the velocity scale we assume an
isothermal atmosphere, $\mathrm{d}T/\mathrm{d}z \sim 0$, so $c_{wave}
= \sqrt{R/c_p} \times \sqrt{gH}$. This amounts to assuming that
gravity waves are very fast \citep[on the order of the speed of sound,
$c_{sound} = \sqrt{\gamma_{a} R T_{eq}} = \sqrt{\gamma_{a} g H}$, where
$\gamma_{a}$ is the adiabatic index; cf.][]{heng2012a}.
As a characteristic length scale we choose the planetary radius $a$. We note that another
potential length scale is given by the equatorial Rossby deformation
radius, $L_{Ro}=\sqrt{a c_{wave}/(2\Omega)}$, which is the maximum
distance that equatorial waves can travel poleward under the influence
of rotation. For slowly rotating planets the Rossby radius exceeds the
planetary radius, $L_{Ro}>a$, and equatorial waves can propagate
planet-wide. We estimate that our choice of $a$ as the length scale is
valid for planets with orbital
period $\gtrsim \mathcal{O}(6)$ days\footnote{Assuming $R=R_{N_2}$,
  $c_p=c_{p,N_2}$, $T_{eq}=300$K, and $a=a_{\Earth}$, $L_{Ro} > a$ for
  a planet beyond a 5.8-day orbital period.}. From conservation of mass, we choose a
vertical velocity scale $c_{wave} \times p_s/a$. The remaining scales
and nondimensionalized equations are shown in Appendix
\ref{sec:appendix1}.

We arrive at the following six nondimensional parameters: \\
\begin{displaymath}
\begin{split}
  {\textstyle
  \left( \frac{R}{c_p}, \frac{2 \Omega a }{c_{wave}},
    \frac{a}{c_{wave}} \frac{g \sigma_{SB} T_{eq}^3}{c_p
      p_s},\frac{\tau_{SW}}{\tau_{LW}},\frac{\kappa_{LW} p_0}{g}
    \left[\frac{p_s}{p_0}\right]^2,
    \frac{C_D a g}{R T_{eq}}
  \right) 
  } \equiv \\
  \left( \frac{R}{c_p}, \frac{a^2}{L_{Ro}^2},
    \frac{t_{wave}}{t_{rad}},\gamma,\tau_{LW}, C_D \frac{a}{H} \right).
\end{split}
\end{displaymath}
The six parameters are related to physical processes as follows:
  the adiabatic coefficient $R/c_p$ controls the lapse rate and is also identical to the
  ratio $2/(2+n)$, where $n$ is the degrees of freedom
  of a gas \citep{pierrehumbert2011b}.
  The nondimensional Rossby radius $a^2/L_{Ro}^2$ governs the
  latitudinal extent over which equatorial waves can transport energy
  and momentum \citep{matsuno1966,showman2011,leconte2013}.
  We emphasize that instead of $a^2/L_{Ro}^2$ one could choose
  different scales and arrive at, for example, a Rossby number or a nondimensionalized
  Rhines scale \citep{showman2010,showman2013b}. Which scale to choose depends on
  the processes under consideration, and our anticipation of
  wave adjustment processes naturally leads to $a^2/L_{Ro}^2$.
  Our results support our analysis, and we find that phase
  curves are largely insensitive to planetary rotation when
  $a^2/L_{Ro}^2 \lesssim 1$, that is, as long as waves propagate
  planet-wide (Section \ref{sec:results}).
  The ratio $t_{wave}/t_{rad}$ compares the
  time it takes for waves to redistribute energy across the planet,
  $t_{wave} \equiv a/c_{wave}$, to the atmosphere's radiative cooling
  time, $t_{rad} \equiv c_p p_s/(g \sigma_{SB} T_{eq}^3)$
  \citep{showman2013b,perez-becker2013a}. The atmospheric
  shortwave and longwave optical thicknesses at the surface are $\tau_{SW} \equiv \kappa_{SW} p_0/g
  \times (p_s/p_0)^2$ and $\tau_{LW} \equiv \kappa_{LW} p_0/g
  \times (p_s/p_0)^2$, and their ratio is $\gamma$. 
  We note that the precise forms of $\tau_{SW}$
  and $\tau_{LW}$ depend on details such as pressure broadening and scattering \citep[e.g.,][]{pierrehumbert2011b,heng2014c}. Our
  definition of $\gamma$ is equal to the more commonly used ratio
  of shortwave to longwave opacities \citep[e.g.,][]{guillot2010} when shortwave
  and longwave opacities have the same pressure dependency.
  The influence of surface friction and surface
  heating on the atmosphere is governed by $C_D a/H$.

Two atmospheres governed by the equations that we assume are
guaranteed to be dynamically similar (identical dynamics in
statistical equilibrium) if their six nondimensional parameters are
identical (also see Section \ref{sec:discussion}). We note that only
the nondimensionalized dynamics will be similar; the physical values
of, for example, temperature gradients or wind speeds could be quite
different. We also note that dynamical similarity does not depend on
how we nondimensionalize the equations, that is, our particular choice
of characteristic scales and nondimensional parameters.
Nondimensionalization therefore allows us to identify atmospheres that
one might consider distinct based on their dimensional parameters, but
that turn out to be dynamically similar.

We test this idea in an idealized GCM. The model is based on the GFDL
Flexible Model System \citep[FMS; ][]{held1994} and was subsequently
modified by \citet{frierson2006} and \citet{merlis2010}. This model
has already been used to simulate the atmospheres of Earth
\citep{frierson2006}, Jupiter \citep{liu2011}, hot Jupiters
\citep{heng2011}, tidally locked terrestrial planets
\citep{merlis2010,mills2013}, and non-synchronously rotating
terrestrial planets \citep{kaspi2014}. For our simulations, we remove
moisture and replace the model's convective parametrization with an
instantaneous dry convection scheme \citep{manabe1965}.
We run all simulations for at least 1000 days with a spatial resolution
of either $64\times128\times30$ or $48\times96\times20$ grid points
(latitude$\times$longitude$\times$vertical, corresponding to T42 or
T31 spectral resolution). We use model time-steps between 30 and 1200 seconds. We vary
  the time-step because hot atmospheres require smaller time-steps for
  numerical stability, whereas colder atmospheres can
  be integrated using longer time-steps but also take longer to reach equilibrium. We expect this behavior,
  given that $c_{wave} \propto \sqrt{T}$, so that hotter atmospheres are more likely to violate the
  Courant-Friedrichs-Lewy criterion.
We consider a simulation
equilibrated once the global-averaged radiative imbalance between incoming
stellar and outgoing longwave radiation has fallen below at least 1\% of
the incoming stellar radiation.
\begin{figure*}[t]
\begin{centering}
% \plottwo{coords_earth.eps}{coords_TL.eps}
% \plotone{EarthVsTL_Dry50_tsurf02.eps}
\includegraphics[width=0.35\textwidth]{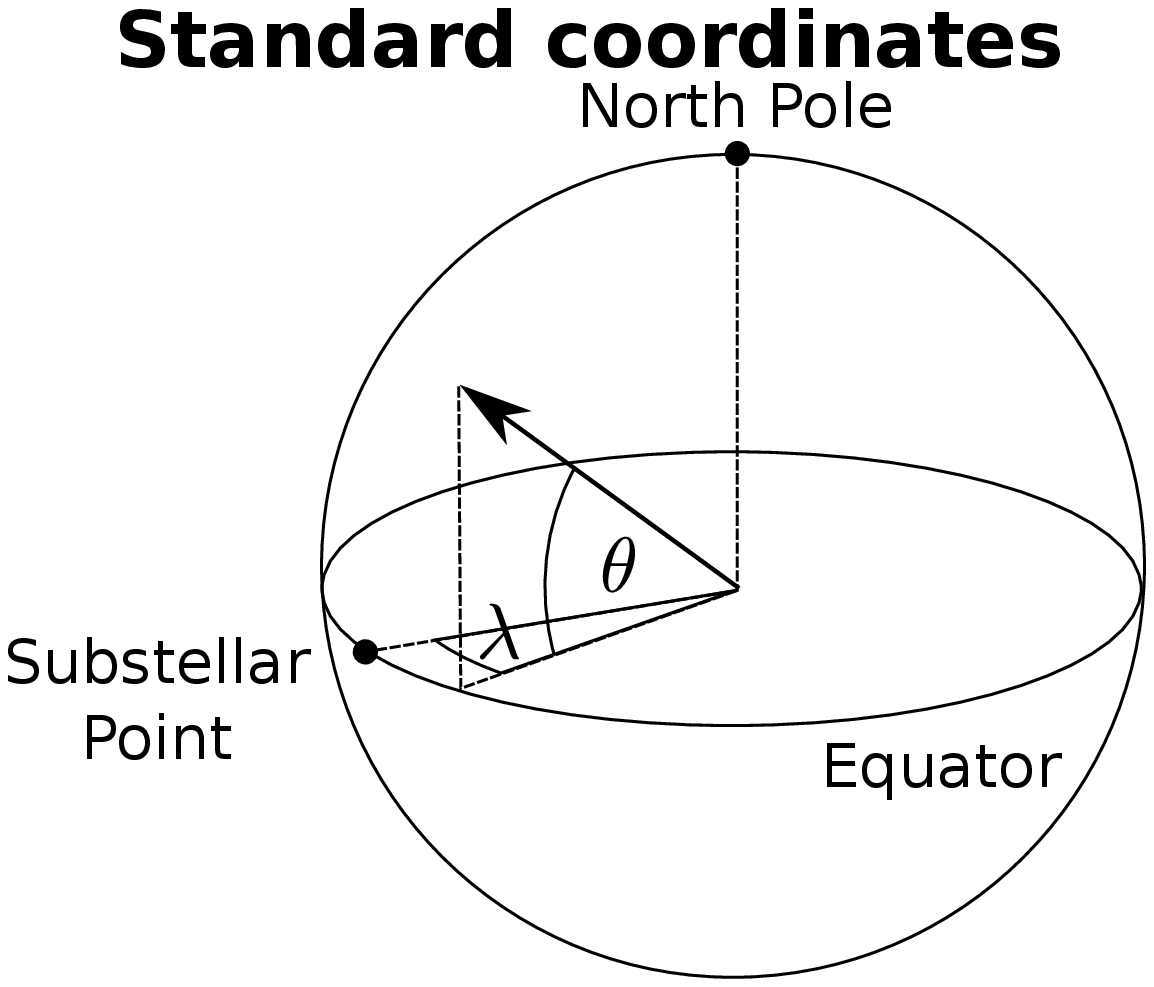}
\includegraphics[width=0.35\textwidth]{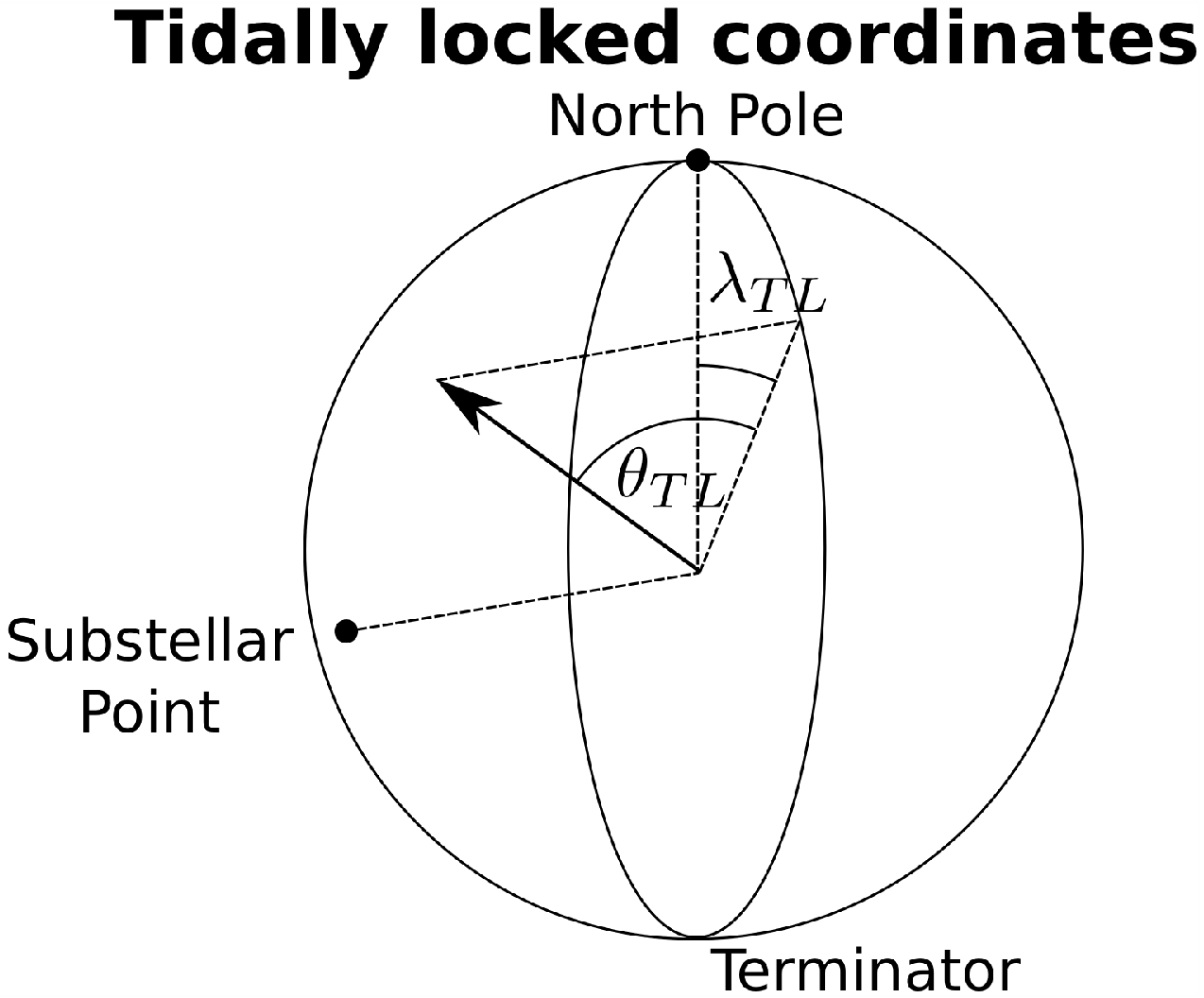}\\
\includegraphics[width=0.7\textwidth]{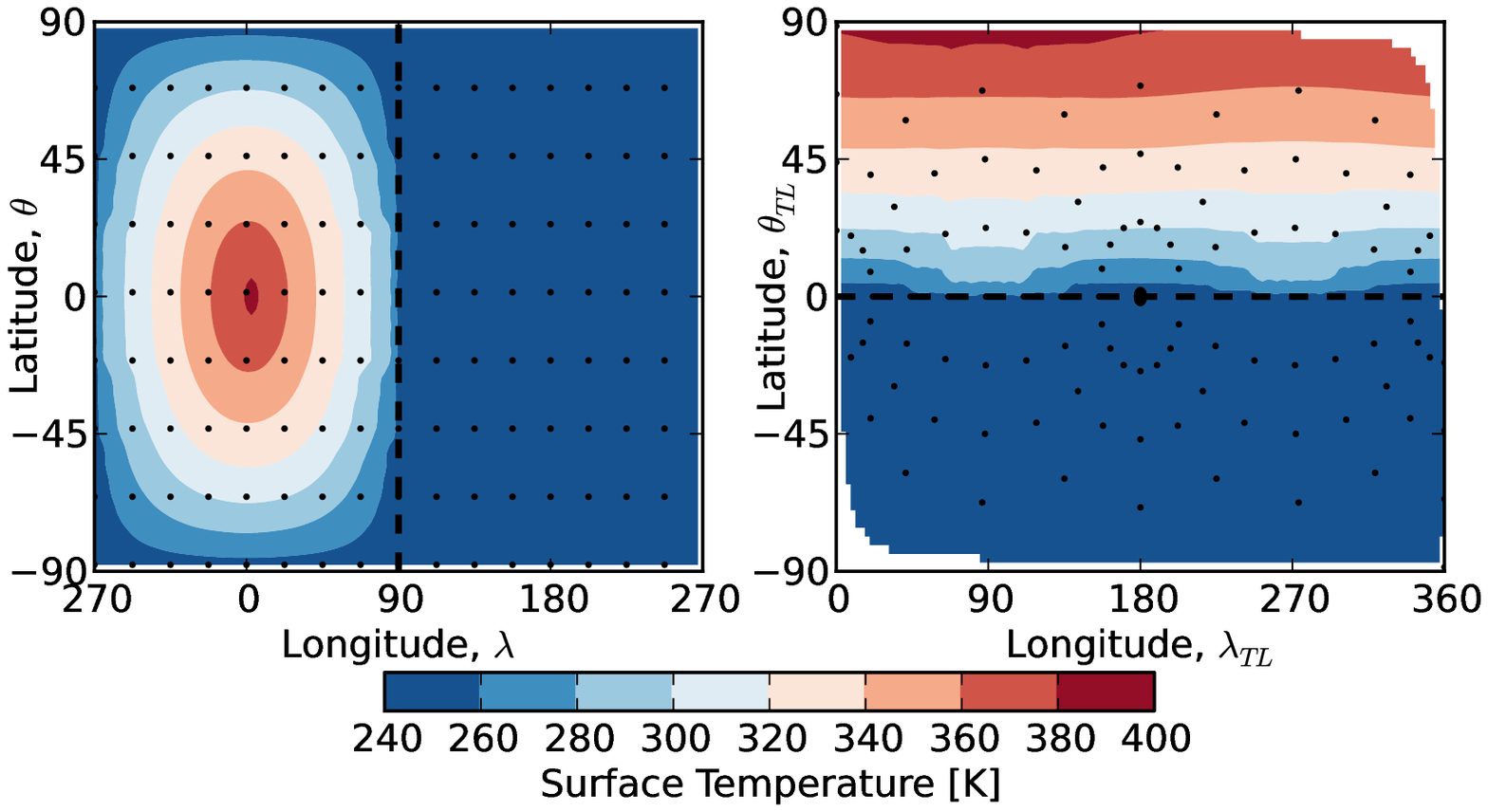}
\caption{Slowly rotating tidally locked planets are
  approximately symmetric about the substellar point. Surface temperature from the reference
  simulation in Table \ref{tab:params} is shown in two different
  coordinate systems. The black dashed line is the
  terminator. Left: In standard coordinates, longitude $\lambda >
  90^\circ$ corresponds to the nightside and the substellar point is
  located at latitude/longitude $(\theta,\lambda)=(0^\circ,0^\circ)$.
  Right: In tidally locked coordinates, tidally locked latitude $\theta_{TL} < 0^\circ$
  corresponds to the nightside and the substellar point is located at
  tidally locked latitude $\theta_{TL}=90^\circ$ (see Appendix \ref{sec:appendix2}). For
  illustration, black dots mark every 64th GCM grid point.}
\label{fig:zero}
\end{centering}
\end{figure*}
We note that the GCM simulates additional higher-order physics, and
therefore contains additional parameters, which we did not include in
the derivation of the six nondimensional parameters. In particular,
the model contains a full Monin-Obukhov surface boundary layer scheme
which self-consistently computes the depth of the boundary layer,
diffusion of surface fluxes, and surface drag. This means the drag
coefficient, $C_D$, is computed by the model instead of being a fixed
parameter\footnote{For the same reason the values for $C_D a/H$ shown
  in our results are only approximate. We estimate $C_D a/H$ assuming
  neutral stratification and $z=10$m. }. For example, for a neutrally
stratified boundary layer $C_D = (k_{vk} / \log[z/z_0] )^2$, where
$k_{vk}$ is the von Karman constant, $z$ is the height of the lowest
model layer and $z_0$ is the roughness length. Because $z$ and $z_0$
only enter into this equation logarithmically we modify $C_D$ by
adjusting $k_{vk}$. Similarly, the GCM requires additional parameters
for its numerical algorithms.
For example, the momentum equations are implemented using numerical dissipation
  via horizontal $\nabla^8$ hyperdiffusion. The hyperdiffusivity is
  chosen to damp the smallest resolved scale on a time scale of 12
  hours, which sets a dissipative timescale. The dynamical core also
  uses a Robert-Asselin time filter, which is controlled by another nondimensional parameter.
Our assumption, which
we test in Section \ref{sec:results}, is that the equations described in
Appendix \ref{sec:appendix1} capture the most important physics
simulated by the GCM.

Although the GCM uses standard latitude-longitude coordinates, we adopt a
tidally locked coordinate system to present our
numerical results. Tidally locked planets in relatively long-period
orbits tend to exhibit a strong symmetry about the axis connecting the
substellar and antistellar points. Figure \ref{fig:zero} shows the
surface temperature in our first reference simulation, which is a
cool, slowly rotating, Earth-sized planet with
$(a,\Omega,T_{eq})=(a_\Earth,2\pi/[50 \mathrm{days}],283 K)$ (see
Table \ref{tab:params}). The surface temperature is symmetric because
the reference simulation is in a slowly rotating dynamical regime,
$a^2/L_{Ro}^2 \ll 1$. The symmetry is not perfect, but it captures the
dominant spatial variability. We therefore define a tidally locked
coordinate system with a tidally locked latitude $\theta_{TL}$ and
longitude $\lambda_{TL}$, where $\theta_{TL}$ is the angle away from
the terminator and $\lambda_{TL}$ is the angle about the substellar
point (see Fig.~\ref{fig:zero}b, Appendix \ref{sec:appendix2}).

To compute phase curves we follow \citet{cowan2008} and assume an edge-on
viewing geometry (see Appendix \ref{sec:appendix3}). We normalize the
disk-integrated fluxes by $F_{rock} \equiv 2/3 \times
L_*(1-\alpha)$, which is the dayside-averaged observer-projected
flux emitted by a planet without an atmosphere. A bare rock will
therefore have a nondimensional phase curve, $F/F_{rock}$, that varies
between zero and one. On the other hand, a planet with efficient heat
transport will have a constant phase curve equal to\footnote{We note that the ratio is not $1/2$. It would be $1/2$
  if we were comparing only dayside-averaged fluxes. The ratio is less here
because we have to additionally account for the observer-projected viewing geometry,
i.e., hotter regions closer to the substellar point appear more prominent to the observer.}
$F/F_{rock} = \sigma_{SB} T_{eq}^4/F_{rock} = 3/8$.
We also define the phase curve peak-to-trough amplitude as the normalized difference
between the phase curve maximum and minimum, $(F_{max}-F_{min})/F_{rock}$.

\section{Sensitivity of phase curves to nondimensional parameters}
\label{sec:results}

\begin{figure*}[!t]
\begin{centering}
%\epsscale{.5}
% \plotone{Multiplot01.eps}
\includegraphics[width=0.8\textwidth]{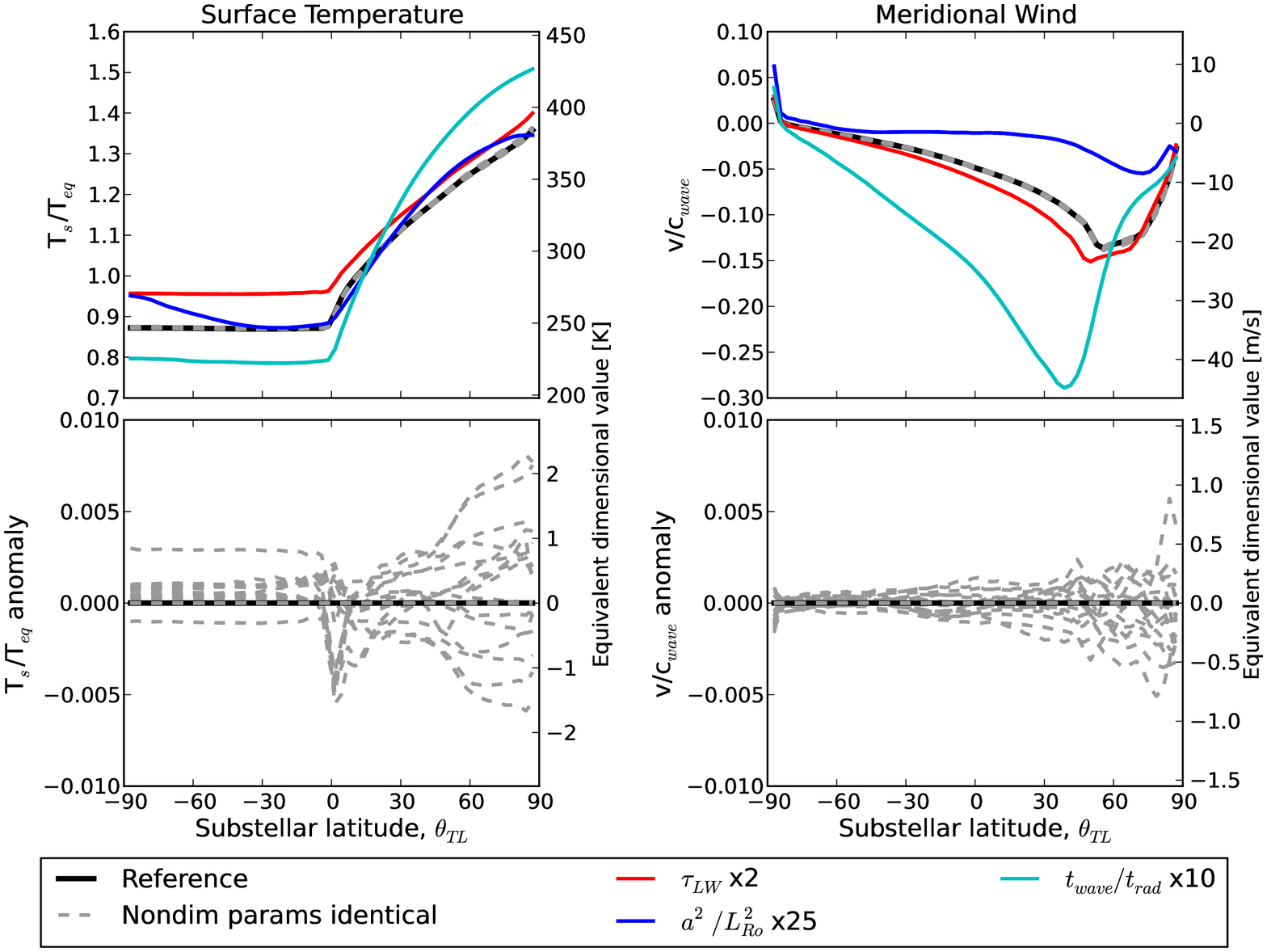}
\caption{Atmospheres have identical dynamics if their
  nondimensional parameters are identical. Top row: Colored curves show simulations in which
we vary the nondimensional parameters. In contrast,
gray dashed curves are simulations in which dimensional parameters are varied,
but nondimensional parameters stay fixed (see Table \ref{tab:params}a). 
Bottom row: the deviation from the
reference when nondimensional parameters stay fixed is $\leq$ 1\% for surface temperature (left)
and generally $\lesssim 3 \%$ for wind velocities (right). Notice the
difference in y-range between the top and bottom row. In addition, plots are
shown in tidally locked coordinates and all
quantities are averaged over tidally locked longitude (see Fig.~\ref{fig:zero}). The meridional
wind is given by the mass-weighted vertical average of meridional wind
between $0.15 \leq p/p_s \leq 0.5$. Wind velocities are negative because
the flow is away from the substellar point.}
\label{fig:one}
\end{centering}
\end{figure*}

First, we test whether our model (with six nondimensional parameters)
captures the main physics of dry, tidally locked atmospheres. We
consider our model adequate if different GCM simulations produce
identical climates when their nondimensional parameters are
identical. Our reference case is a cool, slowly rotating, Earth-sized
planet (Table \ref{tab:params}).  Figure \ref{fig:one} compares the
reference case to simulations in which we change dimensional
parameters, but keep the six nondimensional parameters fixed
(parameter choices are shown in Table \ref{tab:params}a). We find that
the nondimensional surface temperature, $T_s / T_{eq}$, differs less
than $1 \%$ between the reference simulation and the simulations with
fixed nondimensional parameters. The nondimensional meridional wind
velocity, $v/c_{wave}$, in the upper troposphere is more variable, and
differs by $\lesssim 3 \%$ over most of the model domain. The largest
deviation in meridional wind is $\sim 30\%$ near the antistellar
point.  The deviation partly arises because we project simulated wind
speeds into a tidally locked coordinate system, which mixes the wind
vector components. If we instead consider the total wind speed,
$\sqrt{u^2+v^2}/c_{wave}$, the deviation is $<10\%$. Moreover, $v
\rightarrow 0$ at the antistellar point, requiring longer averaging
periods, and wave breaking occurs on the nightside, creating
small-scale structure and numerical dissipation. Both effects can lead
to deviations from dynamical similarity.  For comparison we show some
simulations in which we vary the nondimensional parameters (colored
curves in Figure \ref{fig:one}).  In these simulations surface
temperature and wind velocities change up to $300\%$ compared with the
reference, which demonstrates that the dynamical similarity predicted
by the nondimensionalization is not trivial. We conclude that the
above six nondimensional parameters are sufficient to capture the most
important dynamics of the idealized GCM simulations.

\begin{table*}[!p]
\footnotesize
\begin{tabular}{ l | l }
Dimensional parameters & Nondimensional parameters \\
\tableline
\textit{Reference:} \\
$a = a_{\Earth}$, $\Omega=2\pi/(50~\mathrm{days})$,
$T_{eq}=283$K, $R=R_{N_2}$, $c_p=c_{p,N_2}$, & $\left( \frac{R}{c_p}, a^2/L_{Ro}^2,
    \frac{t_{wave}}{t_{rad}},\gamma,\tau_{LW},   C_D \frac{a}{H}
  \right)$ \\
$p_s$=1 bar, $g=10$ m s$^{-2}$, $\kappa_{LW}=10^{-4}$ m$^2$ kg$^{-1}$,
$\kappa_{SW}$=0 &  $=\left( 0.29, 0.12, 5.1\times10^{-3}, 0, 1, 1.4 \right)$\\
& \\
\tableline
\multicolumn{2}{p{\textwidth}}{
  (a) In 16 simulations we vary the dimensional parameters
  while keeping all nondimensional parameters fixed. For comparison, in
  the bottom three simulations we allow the nondimensional parameters
  to change. \vspace{0.15cm}} \\
\textit{(below: relative to reference)} & \\
$(R,c_p,C_D) \times 2, \Omega \times 2^{1/2}, p_s \times
2^{-3/2}, \kappa_{LW} \times 2^3$ & same as reference \\
$(R,c_p,C_D) \times 2^{-1}, \Omega \times 2^{-1/2}, p_s \times
2^{3/2}, \kappa_{LW} \times 2^{-3}$ & \nodata \\
$(R,c_p,g) \times 2, \Omega \times 2^{1/2}, p_s \times
2^{-1/2}, \kappa_{LW} \times 2^2$ & \nodata \\
$(R,c_p,g) \times 2^{-1}, \Omega \times 2^{-1/2}, p_s \times
2^{1/2}, \kappa_{LW} \times 2^{-2}$ & \nodata \\
$(R,c_p,g) \times 5, \Omega \times 5^{1/2}, p_s \times
5^{-1/2}, \kappa_{LW} \times 5^{2}$ & \nodata \\
$(R,c_p,g) \times 5^{-1}, \Omega \times 5^{-1/2}, p_s \times
5^{1/2}, \kappa_{LW} \times 5^{-2}$ & \nodata \\
$(\Omega,g,\kappa_{LW}) \times 2, a \times 2^{-1}$ & \nodata \\
$(\Omega,g,\kappa_{LW}) \times 2^{-1}, a \times 2^{1}$ & \nodata \\
$\Omega \times (\frac{5}{4})^{1/2}, (T_{eq},g) \times \frac{5}{4}, p_s \times (\frac{5}{4})^{7/2}, \kappa_{LW} \times (\frac{5}{4})^{6}$ & \nodata \\
$\Omega \times (\frac{4}{5})^{1/2}, (T_{eq},g) \times \frac{4}{5}, p_s \times (\frac{4}{5})^{7/2}, \kappa_{LW} \times (\frac{4}{5})^{6}$ & \nodata \\
$(a,R,c_p) \times \frac{3}{2}, (\Omega,p_s) \times (\frac{3}{2})^{-1/2}, \kappa_{LW} \times \frac{3}{2}$ & \nodata \\
$(a,R,c_p) \times 2, (\Omega,p_s) \times 2^{-1/2}, \kappa_{LW} \times 2$ & \nodata \\
$(a,R,c_p) \times \frac{2}{3}, (\Omega,p_s) \times (\frac{2}{3})^{-1/2}, \kappa_{LW} \times \frac{2}{3}$ & \nodata \\
$(a,R,c_p) \times \frac{1}{2}, (\Omega,p_s) \times (\frac{1}{2})^{-1/2}, \kappa_{LW} \times \frac{1}{2}$ & \nodata \\
$(R,c_p,C_D) \times 5, \Omega \times 5^{1/2}, p_s\times 5^{-3/2}, \kappa_{LW} \times 5^3$ & \nodata \\
$(R,c_p,C_D) \times \frac{1}{5}, \Omega \times (\frac{1}{5})^{1/2},
p_s\times (\frac{1}{5})^{-3/2}, \kappa_{LW} \times (\frac{1}{5})^3$ & \nodata \\
\\
$\kappa_{LW} \times 2$ & $\tau_{LW} \times 2$ \\
$\Omega = 2\pi/(2~\mathrm{days})$ & $a^2/L_{Ro}^2\times 25$ \\
$p_s / 10$, $\kappa_{LW} \times 10^2$ & $t_{wave}/t_{rad} \times 10$ \\
\tableline
\multicolumn{2}{p{\textwidth}}{
  (b) This illustrates how we
  vary one nondimensional parameter at a time, while keeping other
  nondimensional parameters fixed (also see Appendix
  \ref{sec:appendix4}). The dimensional parameters remain 
  within the constraints shown in Table \ref{tab:minmax}. \vspace{0.15cm}} \\
\textit{(below: relative to reference)} & \\
$c_p \times 1.5, (R,g) \times 1.5^{1/2}, p_s \times 1.5^{-1/2},
\kappa_{LW} \times 1.5^{3/2}$ & $R/c_p \times 0.82$\\
$(R,c_p,C_D) \times 0.75, p_s \times 0.75^{-3/2}, \kappa_{LW} \times 0.75^{3}$ & $a^2/L_{Ro}^2 \times 1.15$\\
$(R,c_p,C_D) \times 4.64, p_s \times 4.64^{-3/2}, \kappa_{LW} \times 100$ & $a^2/L_{Ro}^2 \times 0.46$\\
$p_s \times 6.32^{-1}, g\times2.5, C_D\times 2.5^{-1},\kappa_{LW} \times 100$ & $t_{wave}/t_{rad} \times 15.8$\\
$p_s \times 0.1, \kappa_{LW} \times 100$ & $t_{wave}/t_{rad} \times 10$\\
$p_s \times 0.141, \kappa_{LW} \times 50$ & $t_{wave}/t_{rad} \times 7.1$\\
$p_s \times 0.5, \kappa_{LW} \times 4$ & $t_{wave}/t_{rad} \times 2$\\
$p_s \times 2^{1/2}, \kappa_{LW} \times 0.5$ & $t_{wave}/t_{rad} \times 0.71$\\
$p_s \times 3.162, \kappa_{LW} \times 0.1$ & $t_{wave}/t_{rad} \times 0.32$\\
$p_s \times 2, g\times2.5^{-1}, C_D\times 2.5,\kappa_{LW} \times 0.1$ & $t_{wave}/t_{rad} \times 0.2$\\
$\kappa_{SW} = 5\times10^{-5}$ & $\gamma = 0.5$\\
$\kappa_{SW} = 10^{-4}$ & $\gamma = 1$\\
$\kappa_{LW} \times 100$ & $\tau_{LW} \times 100$\\
$\kappa_{LW} \times 50$ & $\tau_{LW} \times 50$\\
$\kappa_{LW} \times 10$ & $\tau_{LW} \times 10$\\
$\kappa_{LW} \times 0.5$ & $\tau_{LW} \times 0.5$\\
$\kappa_{LW} \times 0.2$ & $\tau_{LW} \times 0.2$\\
$\kappa_{LW} \times 0.1$ & $\tau_{LW} \times 0.1$\\
$C_D \times 10, (g,p_s) \times 2.5, \kappa_{LW} \times 0.4$ & $C_D a/H
\times 25$\\
$C_D \times 10$ & $C_D a/H \times 10$\\
$C_D \times 0.1$ & $C_D a/H \times 0.1$\\
$C_D \times 0.1, (g,p_s) \times 2.5^{-1}, \kappa_{LW} \times 2.5$ & $C_D a/H
\times 0.04$\\
\tableline
\end{tabular}
\caption{(a) Parameters for the simulations in Figure
  \ref{fig:one}. (b) Parameters for the simulations in Figures
  \ref{fig:two} and \ref{fig:three}a. The drag coefficient $C_D$ is not a fixed parameter in
  the model, so the values shown for $C_D \frac{a}{H}$ is only
  approximate. Symbols for dimensional
  parameters are defined in Table 2, nondimensional parameters
  are defined in Section \ref{sec:methods}.\label{tab:params}}
\end{table*}
%% tab:params
%% tab:dimparams01

Next, we explore how sensitive phase curves are to each of the
nondimensional parameters.  We consider different reference
simulations and vary their nondimensional parameters one at a time to
see how this affects the resulting phase curves. For the reference
simulations we consider different scenarios where $a, \Omega$, and
$T_{eq}$ are fixed.  We do so because $a, \Omega$, and
$T_{eq}$ are relatively easily constrained for a transiting planet.
To vary one nondimensional parameter at a time we
  first find all possible transformations of the dimensional parameters that
  only modify a given nondimensional parameter (see Appendix
  \ref{sec:appendix4}). For example, to decrease $t_{wave}/t_{rad}$ we
  could increase $p_s$ but adjust $\kappa_{SW}$ and $\kappa_{LW}$ such
  that $\gamma$ and $\tau_{LW}$ remain constant. We then use these
  transformations to vary each nondimensional parameter over its
  largest range compatible with fixed $(a,\Omega,T_{eq})$ and the
  constraints in Table \ref{tab:minmax}.

\begin{table*}[pbt]
\begin{centering}
  \begin{tabular}{l c c c c}
      Dimensional Parameter & Symbol & Unit & Minimum value &  Maximum value \\
\tableline
      Planetary radius & $a$ & $a_{\Earth}$ & $1$ & $2$ \\
      Rotation rate & $\Omega$ & days$^{-1}$ & $2\pi/50$ & $2\pi/2$ \\
      Equilibrium temperature & $T_{eq}$ & K &  $100$ & $600$ \\
      Surface gravity & $g$ & 10 m s$^{-2}$ & $\frac{2}{5} \times (a/a_\Earth)$ & $\frac{5}{2} \times (a/a_\Earth)$ \\
      Specific heat capacity & $c_p$ & J kg$^{-1}$ K$^{-1}$ & $820$ & $14230$  \\
      Specific gas constant & $R$ & J kg$^{-1}$ K$^{-1}$ & $190$ & $4157$ \\
      Surface pressure & $p_s$ & bar & $10^{-2}$ & $10$ \\
      Longwave opacity\tablenotemark{a} & $\kappa_{LW}$ & m$^2$/kg & $10^{-5}$ & $10^{-2}$\\
      Shortwave opacity\tablenotemark{a} & $\kappa_{SW}$ & m$^2$/kg & $0$ & $10^{-2}$ \\
      Surface drag coefficient & $C_D$, via $k_{vk}$ & - & $\times 0.1$ & $\times 10$ \\
\tableline
    \end{tabular}\\
    \tablenotetext{a}{Opacities are defined at a reference pressure of
      $p_0 = 1$ bar.}
    \caption{Maximal range of dimensional values we consider. For a given reference simulation we fix
      $(a,\Omega,T_{eq})$, and change the remaining dimensional parameters such
      that only one nondimensional parameter varies at a time. $C_D$ is not a fixed parameter, so we vary the 
      von-Karman constant $k_{vk}$ to increase and decrease $C_D$ by an order of
      magnitude. We vary $R$ and $c_p$, but require that $R/c_p$
      stays within the range of diatomic and triatomic gases
        \citep[$0.22 \lesssim R/c_p \leq 0.29$; section
        2.3.3,][]{pierrehumbert2011b}. The maximum value of $R$ and $c_p$
        corresponds to H$_2$, the minimum value to CO$_2$. In addition we require that
      shortwave optical depth does not exceed longwave optical
        depth ($\gamma \leq 1$).}
  \label{tab:minmax}
\end{centering}
\end{table*}

\begin{figure*}[pbt]
% \begin{centering}
%\plotone{Dry50_curves.eps}
\centering
\includegraphics[width=0.6\textwidth]{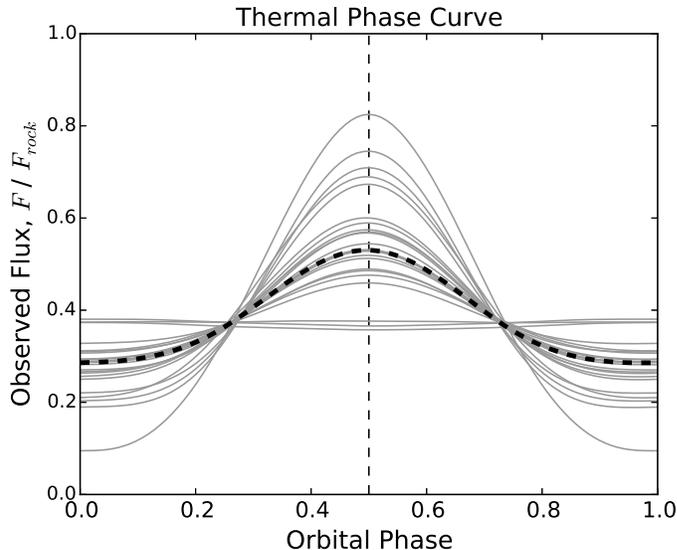}
\caption{For many terrestrial planets, the phase curve's
  peak-to-trough amplitude is sensitive to changes in the atmospheric
  parameters whereas hot/cold spot offsets are small. The dashed black
  line shows the phase curve for the reference simulation in Table
  \ref{tab:params}, and the vertical line indicates secondary
  eclipse. We explore different atmospheric scenarios by varying each
  nondimensional parameter that influences the atmospheric dynamics
  while keeping the other nondimensional parameters fixed
  (Table \ref{tab:params}b).
  The approximately constant curves correspond to
  optically thick atmospheres ($\tau_{LW}\geq10$). The simulations shown here all assume
  $(a,\Omega,T_{eq})=(a_{\Earth}, 2\pi/[50~\mathrm{days}], 283$ K);
  for symbol definitions see Table \ref{tab:minmax}. }
\label{fig:two}
% \end{centering}
\end{figure*}

We find that, for most planets, only the phase curve
peak-to-trough amplitude is robustly sensitive to changes in the nondimensional
parameters.  We start with the above reference scenario of a cool,
slowly rotating planet. The dashed curve in Figure \ref{fig:two} shows
the phase curve of the reference simulation. The planet's thermal flux
is phase-locked with the incoming stellar radiation, that is, there is
no hot spot phase offset. The phase-locking arises because all stellar
radiation is absorbed at the ground ($\gamma=0$), while a significant
part of this energy can also escape directly from the ground to space
without being advected ($\tau_{LW}=1$).
We note that the cold spot offset is larger than the hot spot offset; however, it
would be difficult to detect the cold spot offset because
the phase curve is approximately constant near the antistellar
point (dashed curve in Fig.~\ref{fig:two}).
Next, we vary each nondimensional parameter while keeping the other
nondimensional parameters fixed
(Table \ref{tab:params}b).
The grey lines in Figure \ref{fig:two} show how the phase curve varies
in response to changes in the nondimensional parameters. We find that
the phase curve generally stays phase-locked with the stellar
radiation. This only changes once the atmosphere becomes optically thick
($\tau_{LW} \gg 1$), but in those cases the offset would again be
hard to detect because the planet's thermal emission essentially does
not vary (approximately constant curves in Fig.~\ref{fig:two}).  In
contrast to the negligible hot/cold spot offsets, the phase
curve amplitude is much more sensitive
to changes in the nondimensional parameters (Fig.~\ref{fig:two}).

We explore other reference simulations to see when hot/cold spot offsets become
significant. We find that significant hot/cold spot offsets only occur
when the the atmosphere is optically thick, $\tau_{LW} \gg 1$, the
planet has a high rotation rate, $a^2/L_{Ro}^2 \gtrsim 1$, and is relatively
hot, $t_{wave}/t_{rad} \gtrsim 0.01$. Table \ref{tab:sweep} summarizes our results.
While a cool and slowly rotating planet with $\tau_{LW} = 10$ shows a hot
spot offset of up to $82^\circ$, the phase curve in that case is
almost constant and the offset therefore not detectable
(second-to-bottom row in Table \ref{tab:sweep}). Only in the hottest and most rapidly
rotating scenario with $\tau_{LW} \gg 1$ that we consider do we find a
large hot spot offset of $26^\circ$ which would also be detectable
(bottom row in Table \ref{tab:sweep}).
If future observations found a large hot spot offset, this would
therefore not only imply that the planet has an atmosphere, but also
that the atmosphere would have to be optically thick.
Many terrestrial planets, however, should have small hot spot offsets
(Table \ref{tab:sweep}). For the rest of this paper we therefore focus on
the phase amplitude, and how it could be used to characterize the
atmosphere of a planet.

\begin{sidewaystable*}[p]
  \begin{tabular}{| T | T | T | T | T | T | c | c | c | c | c | c | c  | c | }
    \tableline
    \multicolumn{6}{ | c |}{Reference parameters} & Reference & Ref.~hot &
    \multicolumn{6}{ c | }{Amplitude sensitivity to} \\
    $\frac{R}{c_p}$ & $\frac{a^2}{L_{Ro}^2}$ & $ \frac{t_{wave}}{t_{rad}}$
    & $\gamma$ & $\tau_{LW}$ & $ C_D \frac{a}{H}$ & amplitude & spot offset & $\frac{R}{c_p}$ & $\frac{a^2}{L_{Ro}^2}$ & $ \frac{t_{wave}}{t_{rad}}$
    & $\gamma$ & $\tau_{LW}$ & $ C_D \frac{a}{H}$ \\
% RUN: Dry_50
\tableline
\multirow{2}{*}{0.29} &  \multirow{2}{*}{0.12} &
 \multirow{2}{*}{$5.1\times10^{-3}$} &  \multirow{2}{*}{0} &
 \multirow{2}{*}{1} &  \multirow{2}{*}{1.4} & \multirow{2}{*}{0.24$^{*}$}
 & \multirow{2}{*}{ {\scriptsize$0^\circ$}} & {\scriptsize+0} & {\scriptsize+0} & +0.31 & {\scriptsize+0} & +0.49 & +0.1 \\
& & & & & & &  & {\scriptsize-0.02} & {\scriptsize-0} & {\scriptsize-0.08} & -0.11 & -0.24 & {\scriptsize-0} \\
% RUN: Dry_70
\tableline
\multirow{2}{*}{0.29} &   \multirow{2}{*}{0.3} &
 \multirow{2}{*}{$5.1\times10^{-3}$} &  \multirow{2}{*}{0} &
 \multirow{2}{*}{1} &  \multirow{2}{*}{1.4} & \multirow{2}{*}{0.26}
 & \multirow{2}{*}{ {\scriptsize$3^\circ$}} & {\scriptsize +0} & {\scriptsize +0} & +0.32 & {\scriptsize +0} & +0.48 & {\scriptsize+0.09} \\
& & & & & & &  & {\scriptsize-0.02} & {\scriptsize -0.01} & {\scriptsize-0.07} & -0.11 & -0.25 & {\scriptsize -0} \\
% RUN: Dry_260
\tableline
 \multirow{2}{*}{0.29} &   \multirow{2}{*}{0.6} &
 \multirow{2}{*}{$1\times10^{-2}$} &  \multirow{2}{*}{0} &
 \multirow{2}{*}{1} &  \multirow{2}{*}{2.8} & \multirow{2}{*}{0.36}
 & \multirow{2}{*}{ {\scriptsize$3^\circ$}} & {\scriptsize+0} & {\scriptsize+0.02} & +0.34 & {\scriptsize+0} & +0.42 & {\scriptsize+0.06} \\
& & & & & & &  & {\scriptsize-0.03} & {\scriptsize-0.05} & -0.11 & -0.16 & -0.32 & {\scriptsize-0.02} \\
% RUN: Dry_350
\tableline
 \multirow{2}{*}{0.29} &   \multirow{2}{*}{4.1} &
 \multirow{2}{*}{$6.6\times10^{-2}$} &  \multirow{2}{*}{0} &
 \multirow{2}{*}{1} &  \multirow{2}{*}{1.3} & \multirow{2}{*}{0.8$^{**}$}
 & \multirow{2}{*}{{\scriptsize$7^\circ$}} & {\scriptsize+0} & {\scriptsize+0.01} & +0.16 & {\scriptsize+0} & +0.1 & {\scriptsize+0.09} \\
& & & & & & &  & {\scriptsize-0.03} & -0.16 & -0.34 & -0.13 & -0.51 & -0.1 \\
% RUN: Dry_370
\tableline
 \multirow{2}{*}{0.29} &   \multirow{2}{*}{10} &
 \multirow{2}{*}{$7.6\times10^{-4}$} &  \multirow{2}{*}{0} &
 \multirow{2}{*}{1} &  \multirow{2}{*}{8.1} & \multirow{2}{*}{0.19}
 & \multirow{2}{*}{ {\scriptsize$0^\circ$}} & {\scriptsize+0} & {\scriptsize+0} & +0.32 & {\scriptsize+0} & +0.48 & {\scriptsize+0.04} \\
& & & & & & &  & {\scriptsize-0.03} & {\scriptsize-0.03} & {\scriptsize-0} & {\scriptsize-0.07} & -0.14 & {\scriptsize-0.03} \\
\tableline
% RUN: Dry_640
 \multirow{2}{*}{0.29} &   \multirow{2}{*}{0.12} &
 \multirow{2}{*}{$5.1\times10^{-3}$} &  \multirow{2}{*}{0.5} &
 \multirow{2}{*}{10} &  \multirow{2}{*}{1.4} & \multirow{2}{*}{0.01}
 & \multirow{2}{*}{$82^\circ$} & {\scriptsize+0} & {\scriptsize+0} & {\scriptsize+0} & {\scriptsize+0} & +0.71 & {\scriptsize+0} \\
& & & & & & &  & {\scriptsize-0} & {\scriptsize-0} & {\scriptsize-0.01} & {\scriptsize-0.01} & {\scriptsize-0.01} & {\scriptsize-0} \\
\tableline
% RUN: Dry_660
 \multirow{2}{*}{0.29} & \multirow{2}{*}{4.1} &
 \multirow{2}{*}{$6.6\times10^{-2}$} &  \multirow{2}{*}{0.5} &
 \multirow{2}{*}{10} &  \multirow{2}{*}{1.4} & \multirow{2}{*}{0.5}
 & \multirow{2}{*}{$26^\circ$} & {\scriptsize+0} & {\scriptsize+0.04} & +0.13 & {\scriptsize+0.03} & +0.37 & {\scriptsize+0.06} \\
& & & & & & &  & {\scriptsize-0.05} & -0.17 & -0.31 & {\scriptsize-0} & {\scriptsize-0.01} & {\scriptsize-0.08} \\
\tableline
  \end{tabular}
\caption{We explore a broad range of atmospheric scenarios. The six
  columns on the left show the nondimensional parameters for the
  reference simulations, the center two columns show the phase curve
  amplitude and the hot spot offset in the reference simulation, and
  the six columns on the right show the maximum/minimum change in
  phase curve amplitude in response to each nondimensional
  parameter. The large font emphasizes entries with phase amplitude
  sensitivity bigger than $0.1$. The top row, $*$, corresponds to the cool slowly rotating
  scenario in Figures \ref{fig:two} and \ref{fig:three}a, and the fourth row from
  the top, $**$, corresponds to the hot and rapidly rotating scenario
  in Figure \ref{fig:three}b.}
\label{tab:sweep}
\end{sidewaystable*}

\begin{figure*}[t]
\begin{centering}
% \plottwo{Dry50_deltaF.eps}{Dry350_deltaF_corrected.eps}
\includegraphics[width=0.4\textwidth]{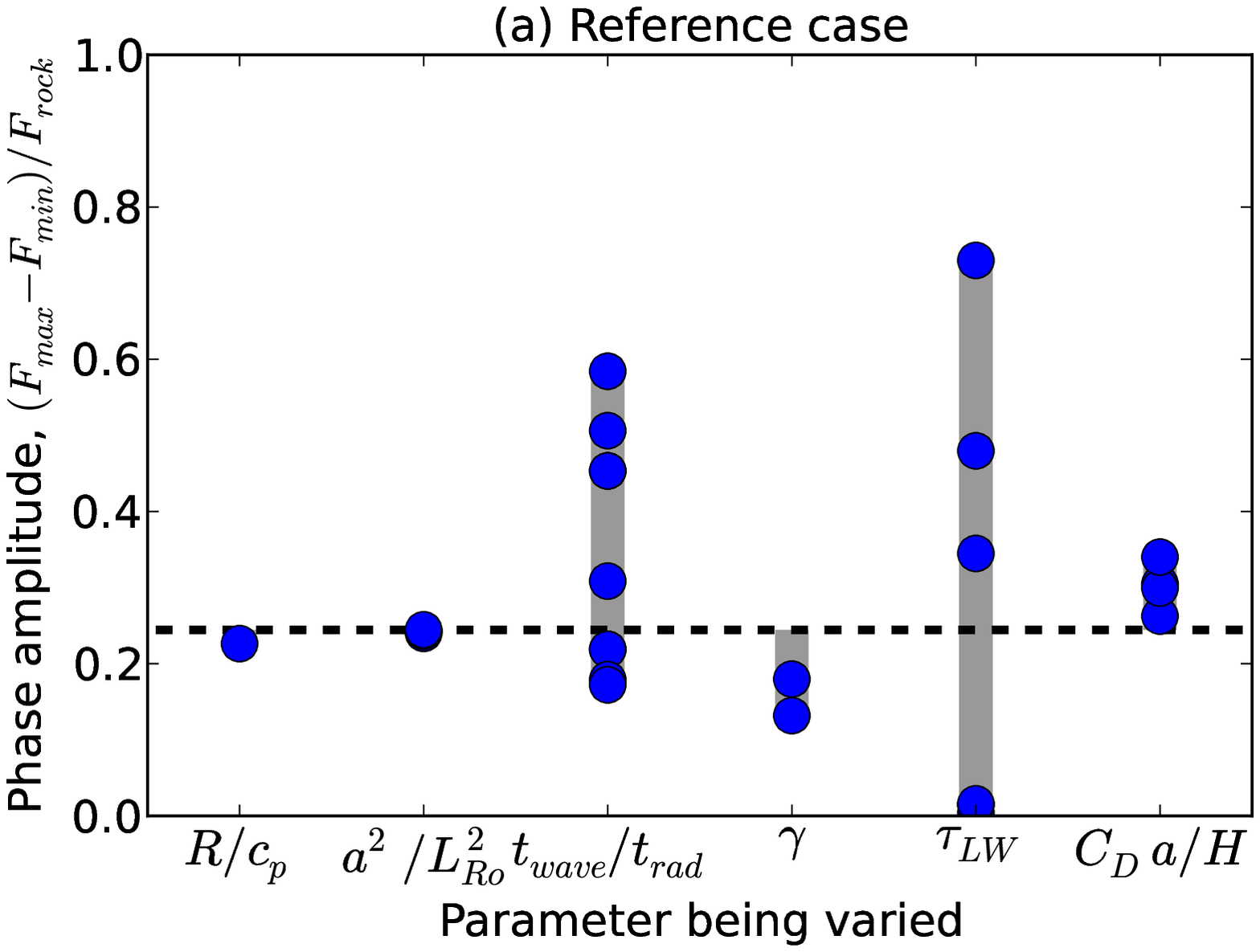}
\includegraphics[width=0.4\textwidth]{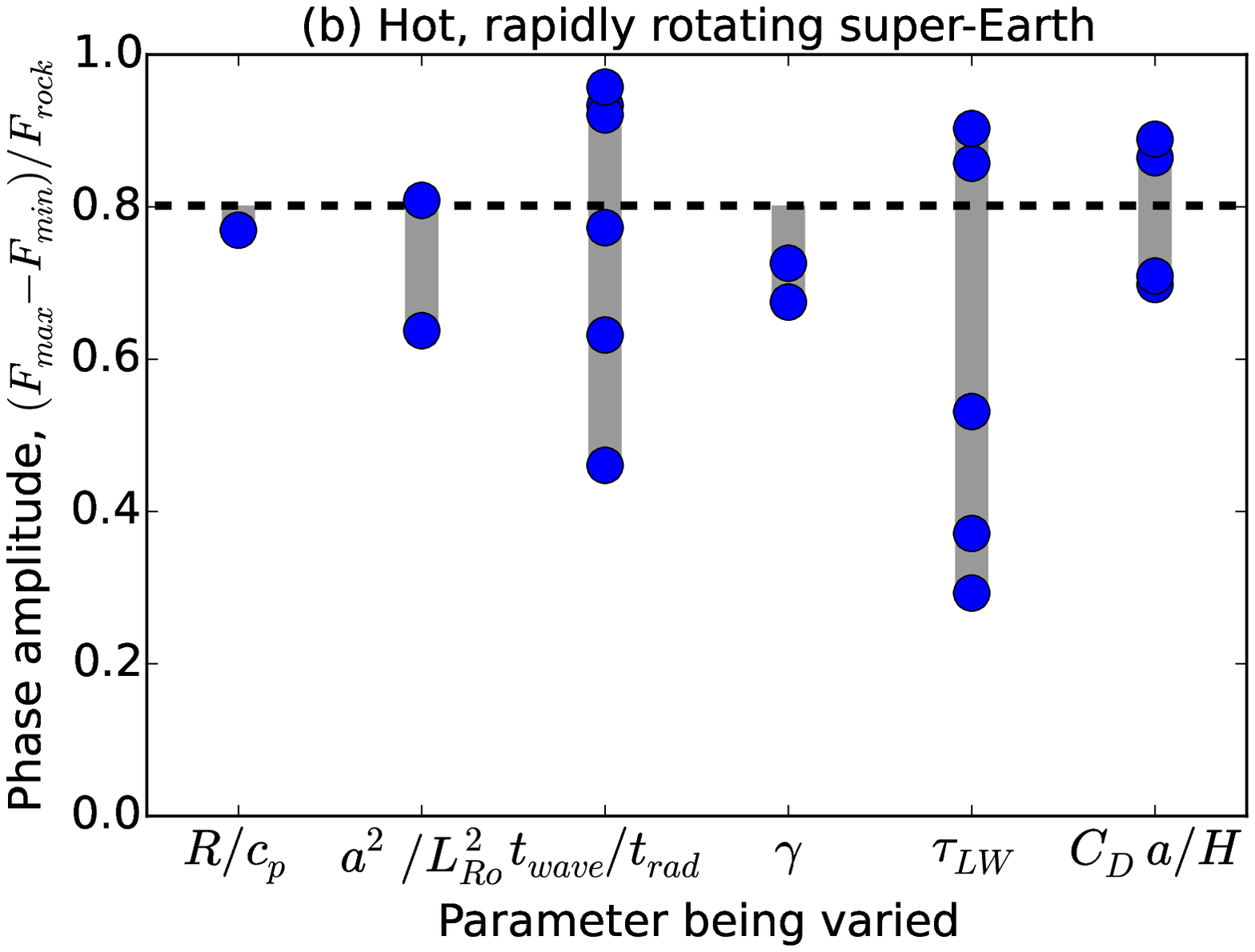}
\caption{Except for hot and
  rapidly rotating planets, the phase curve peak-to-trough amplitude is primarily sensitive to 
  $t_{wave}/t_{rad}$ and $\tau_{LW}$. As in Figure \ref{fig:two}, we vary each nondimensional
  parameter while keeping the other nondimensional parameters
  fixed. The nondimensional parameters are defined in Section
  \ref{sec:methods}. Dashed black lines show the phase amplitude of the
  reference simulations, blue dots show the phase amplitude as nondimensional
  parameters are varied, and vertical bars indicate
  the maximal variation of amplitude, i.e., sensitivity, for each nondimensional parameter. Left: reference
  simulation assumes $(a,\Omega,T_{eq})=(a_{\Earth}, 2\pi/[50~\mathrm{days}], 283$ K).
  Right: reference simulation assumes $(a,\Omega,T_{eq})=(2a_{\Earth},
  2\pi/[2~\mathrm{days}], 600$ K).}
\label{fig:three}
\end{centering}
\end{figure*}

We find that, except for hot and rapidly rotating planets, the phase
amplitude is primarily sensitive to the ratio of dynamical to
radiative timescales, $t_{wave}/t_{rad}$, and the optical depth,
$\tau_{LW}$. We again start with the reference scenario of a cool,
slowly rotating planet. The dashed line in Figure \ref{fig:three}a
shows the reference phase amplitude, $(F_{max}-F_{min})/F_{rock}$, and
the vertical lines show how sensitive the phase amplitude is to
changes in each nondimensional parameter. The phase amplitude is far
more sensitive to $t_{wave}/t_{rad}$ and $\tau_{LW}$ than to any of
the other nondimensional parameters. For example, when we vary
$t_{wave}/t_{rad}$ (primarily by changing $g$ and $p_s$, while
adjusting other parameters; see Table \ref{tab:params}b), the
phase amplitude varies between 0.2 and 0.6. In contrast, when we vary
the nondimensional Rossby radius, $a^2/L_{Ro}^2$ (by changing $R$, and
thus $c_{wave}$, while adjusting other parameters), the phase
amplitude varies by less than 0.01.  We emphasize this does not mean
that the atmospheric dynamics or phase curve are insensitive to
$a^2/L_{Ro}^2$ in general.  As we show in the next paragraph,
$a^2/L_{Ro}^2$ can affect phase curves when $a^2/L_{Ro}^2 \gtrsim 1$.
However, for this particular scenario, once $(a,\Omega,T_{eq})$ are
known then $a^2/L_{Ro}^2$ is already constrained to be much smaller
than one. The remaining observational uncertainty in $a^2/L_{Ro}^2$
barely affects our interpretation of the planet's phase curve
amplitude.  Figure \ref{fig:three}a therefore shows that, for cool,
slowly rotating planets with known $(a,\Omega,T_{eq})$, a phase curve
measurement contains essentially no information about the parameters
$R/c_p$, $a^2/L_{Ro}^2$, $\gamma$, and $C_D a/H$. On the other hand, a
measurement of the phase amplitude would constrain the combination of
$t_{wave}/t_{rad}$ and $\tau_{LW}$.

We explore other reference simulations to determine whether
there are regimes in which the phase amplitude is sensitive to other
parameters. Similar to our result
for hot spot offsets, we find that phase amplitude only becomes
sensitive to $a^2/L_{Ro}^2$, $\gamma$, and $C_D a/H$ for large, hot,
and rapidly rotating planets. Specifically, a planet has to have both
$a^2/L_{Ro}^2 \gtrsim 1$ and $t_{wave}/t_{rad} \gtrsim 0.01$ for
additional nondimensional parameters to affect
the phase amplitude.  Our results are summarized on the right-hand
side of Table \ref{tab:sweep}.  We find that, in all scenarios, the
phase amplitude is most sensitive to $t_{wave}/t_{rad}$ and
$\tau_{LW}$. We also find that $a^2/L_{Ro}^2$ and $t_{wave}/t_{rad}$
both have to be large for the phase amplitude to become sensitive to
additional parameters; a large value of $a^2/L_{Ro}^2$ by itself is
not sufficient (third-to-bottom row in Table
\ref{tab:sweep}).
Together with our above result that hot spot offsets also require
$a^2/L_{Ro}^2 \gtrsim 1$ and $t_{wave}/t_{rad} \gtrsim 0.01$, this suggests that
a regime shift occurs in the atmospheric dynamics near this
threshold.
Figure \ref{fig:three}b shows the scenario in which phase amplitude is
most sensitive to additional parameters (** in Table
\ref{tab:sweep}). This scenario corresponds to a super-Earth with
$(a,\Omega,T_{eq})=(2a_\Earth,2\pi/[2 \mathrm{days}],600 K)$. In this
scenario the phase amplitude is additionally sensitive to variations
in $a^2/L_{Ro}^2$, $\gamma$, and $C_D a/H$ (Fig.~\ref{fig:three}b).
For such a planet, a measurement of the phase amplitude would be
degenerate with multiple atmospheric parameters, although the hot/cold
spot offsets could provide additional information (Table
\ref{tab:sweep}).  Many terrestrial planets, however, will have phase
amplitudes that are, to good approximation, only sensitive to
$t_{wave}/t_{rad}$ and $\tau_{LW}$ (Table \ref{tab:sweep}).

\section{Application to \textit{JWST} observations}
\label{sec:observations}

For planets whose phase curves primarily depend on the ratio of
dynamical to radiative timescales, $t_{wave}/t_{rad}$, and the optical
depth, $\tau_{LW}$, we consider how a phase curve could constrain an
atmosphere's properties.  Expanded in terms of dimensional quantities,
the two nondimensional parameters are $a g \sigma_{SB}
  T_{eq}^{5/2}/(p_s R c_p^{1/2})$ and $\kappa_{LW} p_s^2/(g
  p_0)$. The most important unknowns are the longwave opacity at a
reference pressure, $\kappa_{LW}$, and the surface pressure, $p_s$,
because the other dimensional parameters are relatively easy to
constrain. For a transiting planet, the planetary radius, $a$, would
be known. One can constrain the equilibrium temperature, $T_{eq}$,
because a planet's broadband thermal emission averaged over one orbit
is equal to $\sigma_{SB} T_{eq}^4$. The specific gas constant and heat
capacity, $R$ and $c_p$, vary most significantly between
H$_2$-dominated atmospheres and high mean-molecular-weight
atmospheres. Because $R$ also sets the atmospheric scale height, a
transit spectrum could be sufficient to distinguish between an H$_2$
and a high mean-molecular-weight atmosphere. If one can determine
whether an atmosphere is H$_2$-dominated or not, the detailed value of
$R$ and $c_p$ is secondary; for example, $t_{wave}/t_{rad}$ only
varies by a factor of 2 between a pure N$_2$ atmosphere and a pure
CO$_2$ atmosphere. The surface gravity, $g$, can be constrained via
radial-velocity or transit-timing measurements. Moreover, interior
models indicate that bulk compositions ranging from water ice to iron
would only change the bulk density, and thus $g$, by a factor of
$\sim2$ \citep{seager2007}. In contrast, $\kappa_{LW}$ and $p_s$ can
change the values of $t_{wave}/t_{rad}$ and $\tau_{LW}$ by several
orders of magnitude.

This means a planet's phase amplitude can be used to characterize
longwave opacity, $\kappa_{LW}$, and surface pressure, $p_s$. To
evaluate the feasibility of doing so, we estimate the observable phase
amplitude signal and the precision possible with \textit{JWST}.
For the
signal we assume an optimistic scenario similar to that assumed by \citet{yang2013}.
Specifically, we
assume a cool super-Earth with $(a,T_{eq},\Omega,g,R,c_p) =
(2a_\Earth, 300 K, 2\pi/[10 \mathrm{days}],20 \mathrm{m~s}^{-2},
R_{N_2}, c_{p,N_2})$, orbiting a GJ1214-like star with
$(a_*,T_*)=(0.2 a_\Sun,3000 K)$. We assume the star is $5$ pc away,
and the phase curve is observed between transit and secondary eclipse, for
$5$ days total. 
This planet would have $a^2/L_{Ro}^2 \sim 1$. That is roughly
  the regime for which our findings start to apply, i.e., the planet's
  phase amplitude largely depends on only two nondimensional
  parameters. Cooler and/or smaller planets would be even more
  solidly in the slowly rotating and cool regime, but more difficult
  to observe.
We performed simulations that explore the phase amplitude
as a function of $\kappa_{LW}$ and $p_s$. For a given simulation we
compute the normalized phase amplitude
$(F_{max}-F_{min})/F_{rock}$. We multiply this amplitude by the
planet-star contrast of a bare rock, $F_{rock}/F_*$, to get the phase
amplitude relative to the stellar flux, $(F_{max}-F_{min})/F_*$. To
compute $F_{rock}/F_*$ we approximate the planetary and stellar
emission as blackbody radiation. Following \citet{yang2013}, the planet-star contrast in
some band $[\lambda_1,\lambda_2]$ is then
\begin{displaymath}
  \frac{F_{rock}}{F_*} = \left( \frac{a}{a_*} \right)^2
  \frac{\int_{\lambda_1}^{\lambda_2} B(T_{rock},\lambda) d\lambda}
  {\int_{\lambda_1}^{\lambda_2} B(T_*,\lambda) d\lambda}.
\end{displaymath}
Here $B$ is the Planck function, and $T_{rock}$ is the
dayside-averaged observer-projected temperature of a bare rock
(Appendix \ref{sec:appendix3}). We
assume $16.5 \leq \lambda \leq 19.5~\mu$m, which corresponds to the
F1800W filter on \textit{JWST}'s Mid-Infrared Instrument (MIRI). We
choose this band because it avoids the $15~\mu$m CO$_2$ absorption
feature, but other spectral window regions would be similarly
suitable. For a bare rock the phase amplitude would be
$(F_{max}-F_{min})/F_* = (F_{rock}-0)/F_* = 358$ ppm, while an
atmosphere with perfect day-night heat transport would have a phase
amplitude of $0$ ppm.

\begin{figure*}[!t]
\begin{centering}
% \plotone{N2_nosolar_ppmSignal.eps}
\includegraphics[width=0.6\textwidth]{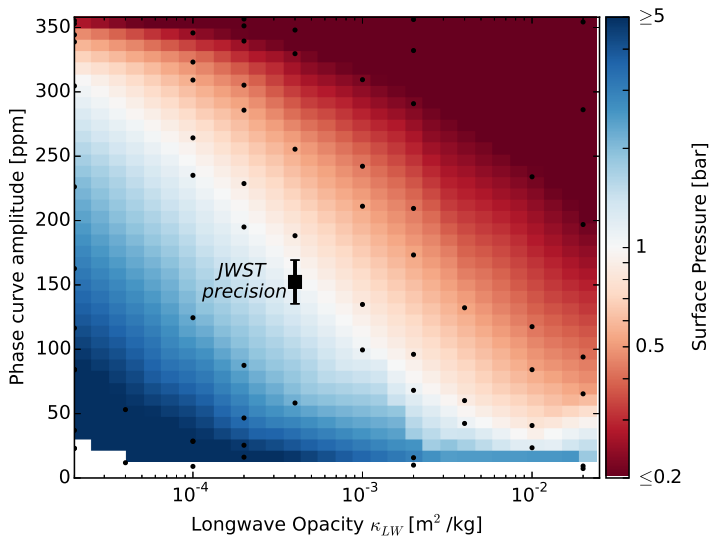}
\caption{This figure illustrates how the observed phase curve amplitude, $[F_{max}-F_{min}]/F_*$,
  (y-axis) can be used to infer the surface pressure of a planet
  (colors).  Black dots indicate simulations, and the color scale is
  interpolated between simulations. A
  given amplitude is compatible with both a thin atmosphere that is
  opaque to longwave radiation (large $\kappa_{LW}$) and a thick
  atmosphere that is transparent to longwave radiation (small
  $\kappa_{LW}$). If $\kappa_{LW}$ is known, for example from transit
  spectroscopy, then the phase curve constrains surface pressure and atmospheric mass. We
  assume a super-Earth around a GJ1214b-like star, such that a bare
  rock would exhibit a phase curve amplitude near $18~\mu$m of $358$
  ppm. The black square shows a representative $1$ bar atmosphere, and
  the error
  bars show our optimistic estimate for \textit{JWST}'s
  $\pm1\sigma$ precision on the phase amplitude (see Section
  \ref{sec:observations}). For
  reference, opacities in Solar System atmospheres tend to fall
  within an order of magnitude of $\kappa_{LW}\sim 10^{-3}$
  m$^2$ kg$^{-1}$ \citep{robinson2014b}.}
\label{fig:four}
\end{centering}
\end{figure*}

To estimate the precision possible with \textit{JWST}, we assume that
the observational error is dominated by stellar photon noise.
In the photon noise limit the precision is
$\sigma/F_* = 1/\sqrt{N}$, where the number of stellar photons $N$ is
\begin{displaymath}
  N = \pi \left( \frac{D}{2} \right)^2 \Delta t \left( \frac{a_*}{d}
  \right)^2 \int_{\lambda_1}^{\lambda_2} \frac{B(T_*,\lambda)}{E(\lambda)} d\lambda.
\end{displaymath}
Here $D$ is the diameter of \textit{JWST}'s mirror (=$6.5$m), $\Delta t$ is the
length of observation, $d$ is the distance between observer and star,
and $E(\lambda)=hc/\lambda$ is the energy per photon.
We first make an optimistic estimate for the precision. Recent measurements with the \textit{Hubble Space Telescope} almost reached the photon noise limit
\citep{kreidberg2014a,knutson2014b}, and we assume
\textit{JWST} will do similarly well. To account for imperfect
instrument throughput and detector efficiency we degrade the photon-limited
precision by a factor of $1/3$ \citep[Fig.~3,][]{glasse2010}. We find
that, over a 12 hour integration, \textit{JWST} should be
able to measure the planet-star flux ratio with a precision of $12$
ppm\footnote{We note that \citet{yang2013}
  similarly estimate \textit{JWST} precisions assuming the photon noise
  limit. However, those calculations contained an error and the resulting
  estimates are too small by a factor of a few (N.~Cowan, personal communication).}. Because the phase
amplitude is the difference between two fluxes, the $3\sigma$ uncertainty interval for
the phase amplitude is\footnote{The factor of $\sqrt{2}$
  assumes that uncertainties between different observation
  periods are uncorrelated. The uncertainty on the phase amplitude is
  then related to the uncertainty of a single observation period
  as $\sigma_{amplitude} = \sqrt{ \sigma_{single}^2 + \sigma_{single}^2} = \sqrt{2}
  \sigma_{single}$.} $\sqrt{2}\times 3 \times 12$ ppm = $51$ ppm.
For a pessimistic estimate we repeat the previous calculation, but additionally impose a noise floor of
$40$ ppm. This floor represents unexpected instrumental
systematics, zodiacal light or other noise sources. With this noise floor, a 12 hour
integration would only reach $10\%$ of the photon noise limit. Given that \textit{Spitzer}
measurements were able to reach $\sim30\%$ of the photon noise
limit \citep[Fig.3,][]{cowan2012a}, we consider this estimate very pessimistic.
In this case a $2\sigma$ ($3\sigma$) measurement of the phase amplitude
would have a precision of $\sqrt{2}\times 2\times40= 113$ ($170$) ppm.

Figure \ref{fig:four} shows our simulation results and optimistic
$1\sigma$ precision estimate.
We find that thin atmospheres ($p_s\leq 0.2$ bar) have
  phase amplitudes close to $358$ ppm for small and moderate values of
  $\kappa_{LW}$  ($\leq 10^{-3}$ m$^2$ kg$^{-1}$). Taking into account
  measurement uncertainties of $\sim50$ ppm, these atmospheres would
  be difficult to distinguish from bare rocks. Similarly, thick
  atmospheres ($p_s > 5$ bar) tend to have phase amplitudes close to zero.
A phase curve would constrain atmospheric mass most
effectively between those two limits.
Figure \ref{fig:four} also shows that, between those limits, the
phase amplitude is sensitive to both $\kappa_{LW}$ and
$p_s$. We find that any observed phase amplitude would be compatible with
both a thin atmosphere that is opaque to longwave radiation (large
$\kappa_{LW}$) and a thick atmosphere that is transparent to
longwave radiation (small $\kappa_{LW}$).
Nevertheless, if transit or emission spectroscopy could determine the
concentration of greenhouse gases in an atmosphere, and therefore
$\kappa_{LW}$, the phase amplitude would yield the value of $p_s$.
As an example we highlight a simulation with a $1$ bar atmosphere and
$\kappa_{LW} = 4\times10^{-4}$ m$^2$ kg$^{-1}$ (black square in
Fig.~\ref{fig:four}). The exact value of this atmosphere's phase
amplitude is $152$ ppm. The observed phase amplitude would therefore
be $152\pm51$ ppm with $3\sigma$ confidence,
which constrains the surface pressure to $0.7 \leq p_s\leq 1.3$
bar. Even using our pessimistic precision
estimate, we find that \textit{JWST} would be able to constrain
surface pressure to $0.5 \leq p_s\leq 2.3$ bar, albeit only with
$2\sigma$ confidence ($152 \pm 113$ ppm). We further note that our pessimistic precision
estimate would only place a lower bound on the surface pressure, $p_s
\geq 0.2$ bar, with $3\sigma$ confidence ($152 \pm 170$ ppm).
These values are the most precise constraints
  that the phase curve amplitude can place on surface pressure, because we
  assumed the other dimensional parameters
  ($\kappa_{LW}$,$R$,$c_p$, etc.)~are already well
  characterized via transit or emission spectroscopy. Observational
  uncertainties in the other dimensional parameters would
  increase the uncertainty in the inferred surface pressure.
Nevertheless, our results show that
a phase amplitude measurement can place meaningful bounds on a planet's atmospheric
mass, while the necessary observation time is competitive with the
time required to constrain atmospheric mass via
transit or emission spectroscopy (cf. Section \ref{sec:introduction}).

\section{Discussion}
\label{sec:discussion}

Dimensional analysis is a crucial tool in comparative planetology and
the study of exoplanets \citep{golitsyn1970,mitchell2010,showman2010,read2011,potter2013,delgenio2013,mitchell2014}.
Our approach highlights the utility of the Buckingham-Pi theorem for
the study of planetary atmospheres \citep[cf.][]{frierson2005}.  We
show that the primitive equations coupled to the two-stream equations
are governed by a fairly small set of nondimensional parameters. These
nondimensional parameters also encapsulate the atmospheric dynamics of
an idealized GCM.  Our analysis reveals basic dimensional
degeneracies, which could allow modelers to sample large parameter
spaces more efficiently (Fig.~\ref{fig:one}, Appendix \ref{sec:appendix4}). 
It is straightforward to expand our
analysis to include additional physics, for example, moist
thermodynamics, multi-band radiation, non-hydrostatic atmospheres,
chemical disequilibrium or magnetohydrodynamics. It follows from the
Buckingham-Pi theorem that each additional independent physical
parameter will introduce another nondimensional parameter.

Our analysis suggests that the dynamics of gaseous planets could be
even easier to understand than the dynamics of terrestrial
planets. Gaseous planets do not have a distinct surface, so we suppose
that their atmospheric dynamics are to first order independent of the
bottom boundary.  A range of modeling studies tend to support this
assumption \citep{heng2011,menou2012a,kataria2013}.  This means the
dynamics are insensitive to the surface friction/heating parameter,
$C_D a/H$. We furthermore need to replace the surface pressure, $p_s$,
with a new characteristic pressure. We note that for terrestrial
planets $p_s$ denotes the depth of the dynamically active part of the
atmosphere, where winds are driven by gradients in the stellar
forcing. For a gaseous planet we analogously use the photon deposition
depth, that is, the pressure where stellar radiation is absorbed,
$p_{D} \sim \sqrt{g p_0/\kappa_{SW}}$ \citep[see
  eqn.~117,][]{heng2014c}. The longwave optical depth $\tau_{LW}$,
which was previously defined at the surface, now becomes the optical
depth at the level of photon deposition, $\tau_{LW} = \kappa_{LW}
p_0/g \times (p_{D}/p_0)^2 = \kappa_{LW}/\kappa_{SW} =
\gamma^{-1}$. The last step shows that $\tau_{LW}$ and $\gamma$ cease
to be independent degrees of freedom. This is also consistent with the
Buckingham-Pi theorem, because we should lose one nondimensional
parameter in the limit $p_s \rightarrow \infty$. The dynamics of
gaseous planets then depend on only four nondimensional parameters
(where $p_s$ is now replaced by $p_{D}$):
\begin{displaymath}
 \left( \frac{R}{c_p}, \frac{a^2}{L_{Ro}^2}, \frac{t_{wave}}{t_{rad}},\gamma \right).
\end{displaymath}
This set is sufficiently small for easy numerical
exploration. Moreover, observations that cannot be explained by a
model with the above four parameters would strongly point to the
importance of additional physics, for example, breakdown of well-mixed
gaseous opacities (via chemical disequilibrium), non-grey radiative
effects, condensation or clouds.

Next, we discuss the observational effort necessary for phase curve
observations. While there were initial attempts to monitor phase
curves discontinuously with \textit{Spitzer} \citep{harrington2006},
more recent \textit{Spitzer} observations tend to cover the whole
course of an orbit to account for long-term instrumental drift
\citep{knutson2012}.  Similar continuous observations with
\textit{JWST} would become very time-consuming for planets with period
$\gtrsim$10 days. Although we assumed in Section \ref{sec:observations} that the planet
is observed for half an orbit, our results
indicate that full phase curve coverage is not necessary as long as
hot spot offsets are small or negligible (Fig.~\ref{fig:two}). In
theory this means that one only needs to observe a planet's thermal
emission near primary and secondary eclipse, which would greatly
reduce the required observation time.
In practice, it will be challenging to relate the observed thermal fluxes from distinct
observation periods, but precise characterization of \textit{JWST}'s
instrumental drift might still permit discontinuous observation strategies.

Finally, we discuss further physics that might influence our
conclusions. Any model necessarily only approximates the dynamics of real
atmospheres. The dynamical similarities predicted by our analysis will
break down if processes neglected in our model become significant
\textit{or} if the neglected processes are singular perturbations. We
do not expect condensation to be a singular perturbation, given that
dry models are able to reproduce many aspects of Earth's
atmospheric dynamics \citep{schneider2006}.
For a planet \textit{with} a condensing substance, our
method should provide an upper bound on the atmospheric mass. That is
because latent heat transport, and for true Earth analogs
ocean heat transport, would increase the day-night energy
transport.
Clouds would similarly reduce the phase curve amplitude,
by reducing the dayside brightness temperature while not strongly
affecting the nightside brightness temperature \citep{yang2013}. If this
effect is strong enough it can even reverse the expected day-night
phase curve pattern. Detection of a inverted phase curve
pattern would therefore be a tell-tale sign of a condensing
atmosphere. In all other cases, a planet with a condensing substance
should have a reduced phase curve amplitude and thus resemble a dry planet
with larger atmospheric mass.

For the radiative transfer, we assume that shortwave and
  longwave opacities both increase linearly with pressure due to
  pressure broadening and collision-induced absorption. This
  assumption breaks down if atmospheric opacities are set by different mechanisms, for
  example, if shortwave radiation was
  absorbed by dust (which is insensitive to pressure broadening). We therefore explored
  simulations in which the shortwave opacity is independent of
  pressure. We find that reducing the pressure dependency has an
  effect qualitatively similar to increasing the ratio of shortwave to
  longwave optical depths, $\gamma$, in our standard
  simulations. Specifically, the resulting increase in shortwave
  absorption at higher altitudes creates stratospheric inversions,
  but only has a limited impact on thermal phase curves ($\sim10-15\%$
  decrease in phase amplitude compared to our standard simulations
  with $\gamma=1$; see Fig.~\ref{fig:three}).

We also assume that planets are tidally locked, but the
effect of atmospheric thermal tides \citep{cunha2014} and
trapping in higher-order resonances \citep{makarov2012a} could result
in non-synchronous orbital states.
Non-synchronous rotation would introduce an additional
  nondimensional parameter into our model that compares the length of
  the day-night cycle with the planet's cooling timescale. We do
  not expect that a small deviation from synchronous rotation would be
  a singular perturbation of this parameter. \citet{yang2014a}
  show that non-synchronously rotating planets with sufficiently slow
  rotation and/or short cooling timescale smoothly approach the tidally
  locked regime (their Fig.~1a). Such planets rotate
  non-synchronously but in an instantaneous sense still
  appear tidally locked.
\citet{cunha2014} show that thermal tides are generally not
compatible with synchronous rotation.
Nevertheless, the deviation from synchronous rotation is
small for planets in close orbits with zero eccentricity around
small stars, such as M-dwarfs \citep[Table 2 in][]{cunha2014}. Unless synchronous rotation
represented a singular limit (see above), the potential
effect of thermal tides should then not greatly affect the types of
planets we consider here.
Many terrestrial planets might never reach synchronous rotation
and instead get trapped in a higher-order orbital resonance, like
Mercury did in our Solar system 
\citep{makarov2012a}. The probability of being trapped decreases for
planets with lower eccentricity \citep[Fig.~6 in][]{makarov2012a}, which means that our
assumption of synchronous rotation is at least consistent with the
fact that also we do not consider non-zero eccentricities.
It is beyond the scope of this article to investigate higher-order
spin states, but phase curves at visible wavelengths could
provide a consistency check for applying our method to planets with
optically thin atmospheres \citep{fujii2014}.

\section{Conclusions}

We use dimensional analysis to find a set of six nondimensional
parameters that captures the main atmospheric dynamics of dry,
tidally locked terrestrial planets in an idealized GCM.
We use the
GCM to investigate the sensitivity of thermal phase curves to each of
the nondimensional parameters. Except for hot and rapidly rotating
atmospheres that are optically thick in the longwave, we do not find
significant hot spot offsets. On the other hand, the phase curve
amplitude remains sensitive to changes in the atmospheric parameters
across a large range of atmospheric scenarios. Focusing on the phase
amplitude, we find that the phase amplitude of many terrestrial
planets is sensitive to only two nondimensional parameters. The main
unknowns in the two nondimensional parameters are the surface pressure
and the longwave opacity. The longwave opacity can be constrained by
transit or emission spectroscopy, in which case the phase amplitude
would constrain the surface pressure and atmospheric mass.  As an
example, we estimate that a broadband phase curve near $18~\mu$m
with \textit{JWST}, taken over a single half-orbit, could be sufficient to
constrain the surface pressure of a cool super-Earth to within a
factor of two. Constraints like the one we propose will be crucial for
understanding atmospheric evolution, in particular atmospheric
escape. Moreover, constraining atmospheric mass is important for
characterizing the surface conditions of potentially habitable
planets.

\acknowledgments

We are grateful to Nick Cowan, Laura Kreidberg, Feng Ding, Wendy Zhang,
and Jacob Bean for their insightful comments on an early draft, and
Kevin Heng for thoughtfully reviewing our work. We
additionally thank Nick Cowan, Laura Kreidberg, and Sean Mills for
discussing \textit{JWST} precision estimates with us.
D.D.B.K. was supported by NSF DMS-0940261, which is part
of the Mathematics and Climate Research Network. D.S.A. acknowledges
support from an Alfred P. Sloan Research Fellowship. This work was
completed with resources provided by the University of Chicago
Research Computing Center.

\appendix

\section{Appendix: Basic equations}
\label{sec:appendix1}

The primitive equations expressed in standard latitude-longitude-pressure
coordinates ($\theta,\lambda,p$) are
\begin{mathletters}
\begin{eqnarray*}
\frac{D\mathbf{u}}{Dt} &=& -2 \Omega \sin \theta \mathbf{k} \times \mathbf{u} - \nabla\phi -g \frac{\partial \mathbf{F}_{m} }{\partial p}, \\
\frac{\partial \phi}{\partial p}&=& -\frac{R T}{p}, \\
\nabla \cdot \mathbf{u} & + & \frac{\partial \omega}{\partial p}= 0, \\
\frac{DT}{Dt}&=& \frac{R T \omega}{c_p p} + \frac{g}{c_p}
\frac{\partial F_{rad} }{\partial p} + \frac{g}{c_p} \frac{\partial
  F_{sens}}{\partial p} + \frac{g}{c_p} \frac{F_{conv}}{\partial p}.
\end{eqnarray*}
\end{mathletters}
Here $\mathbf{u} = (u,v)$ is the horizontal wind velocity,
$\frac{D}{Dt} = \frac{\partial}{\partial t} + \mathbf{u} \cdot
\nabla + \omega \frac{\partial}{\partial p}$ is the material
derivative, $\mathbf{k}$ is a unit vector pointing along the axis of planetary rotation, $\phi$ is
the geopotential, $T$ is temperature, $\omega \equiv \DDt{p}$ is the
vertical velocity (expressed as a change in pressure), and dimensional
parameters are defined in Section \ref{sec:methods}.  From the top,
these equations express conservation of momentum, the hydrostatic
approximation, conservation of mass, and conservation of
energy. Although the mass conservation equation looks like it assumes
incompressibility, it does not. The primitive equations are
compressible, but mass conservation can be written in the above simple
form using $p$ coordinates and the hydrostatic equation
\citep[p.79]{vallis2006}. The forcing terms are
\begin{mathletters}
\begin{eqnarray*}
F_{rad}&=&F_{LW}^{\uparrow}-F_{LW}^{\downarrow}-F_{SW}^{\downarrow}, \\
\frac{\partial F_{LW}^{\uparrow}}{\partial p}&=& \frac{2 \kappa_{lw}}{g} ( F_{LW}^{\uparrow} -\sigma_{SB} T^4), \\
\frac{\partial F_{LW}^{\downarrow}}{\partial p}&=& -\frac{2 \kappa_{lw}}{g} (F_{LW}^{\downarrow}-\sigma_{SB} T^4), \\
\frac{\partial F_{SW}^{\downarrow}}{\partial p} & = &-\frac{2 \kappa_{sw}}{g} F_{SW}^{\downarrow},\\
F_{SW}^{\downarrow}|_{p=0} & = &
 \begin{cases}
 (1-\alpha) L_* \cos \theta \cos \lambda & \mathrm{if~ } 270^\circ
 \leq \lambda \leq 90^\circ \\
 0 & \mathrm{elsewhere},
 \end{cases}\\
\mathbf{F}_{m} |_{p_s}&=& \rho_s C_D |\mathbf{u}_s| \mathbf{u}_s, \\
F_{sens} |_{p_s}&=& \rho_s c_p C_D |\mathbf{u}_s| (T_s-T|_{p_s}).
\end{eqnarray*}
\end{mathletters}
Here $\mathbf{u}_s$ is the surface wind velocity, $T_s$ is the surface
temperature, $T|_{p_s}$ is the near-surface air temperature, and $\rho_s$ is the atmospheric
density at the surface ($\rho_s =  p_s R^{-1} T|_{p_s} $, using the
ideal gas law). The convective heat flux $F_{conv}$ instantaneously 
adjusts an unstable lapse rate toward the dry adiabat while conserving dry enthalpy
\begin{eqnarray*}
  \ddt{} \int_0^{p_s} c_p T \frac{dp}{g} = 0.
\end{eqnarray*}

We neglect scattering in the radiative equations. We assume
  the hemi-isotropic closure, which is why the radiative equations contain a
  factor of two \citep[Section 6.4 in][]{heng2014c}. We also assume that
opacities increase with $p$ due to pressure broadening and/or
  collision-induced absorption, so $\kappa_{sw
}=\kappa_{SW} (p/p_0) $ and $\kappa_{lw }=\kappa_{LW} (p/p_0) $. We define nondimensional shortwave and longwave optical depths as
$d\hat{\tau}_{sw}/dp = 2\kappa_{sw}(p)/g$ and $d\hat{\tau}_{lw}/dp
= 2\kappa_{lw}(p)/g$. We integrate to find the total optical depths,
$\tau_{SW}\equiv \hat{\tau}_{sw}(p_s)=\kappa_{SW} p_0/g \times
(p_s/p_0)^2$ and $\tau_{LW}\equiv \hat{\tau}_{lw}(p_s) = \kappa_{LW}
p_0/g \times (p_s/p_0)^2$.
The surface temperature, $T_s$, is determined by energy balance
\begin{eqnarray*}
  \mathcal{C} \ddt{T_s} & = & F_{SW}^{\downarrow} |_{p_s} - (\sigma_{SB}
  T_s^4 - F_{LW}^{\downarrow}|_{p_s} ) - F_{sens} |_{p_s}.
\end{eqnarray*}
Here $\mathcal{C}$ is the surface thermal inertia. For a tidally
locked planet the stellar forcing does not depend on time. As a first
approximation we ignore internal atmospheric variability and assume
that $T_s$ is time-independent. This means that $\mathcal{C}$ does not
enter the list of dimensional quantities (see below for a discussion
of when this assumption is valid).

We form the following nondimensional quantities, marked with
the hat symbol: $\frac{D}{Dt} =
\frac{1}{t_{wave}}\left(\frac{D}{D\hat{t}}\right)$, $\nabla =
\hat{\nabla}/a$, $p=p_s\hat{p}$, $\mathbf{u} = c_{wave} \hat{\mathbf{u}}$, $\omega = c_{wave}
p_s/a \times \hat{\omega}$, $\phi = g H \hat{\phi}$, $T = T_{eq}
\hat{T}$, $F_{rad} = \sigma_{SB} T_{eq}^4 \hat{F}_{rad}$, $F_{conv} =
\sigma_{SB} T_{eq}^4 \hat{F}_{conv}$, $\mathbf{F}_{m} = p_s C_D c_{wave}^2 /(R
T_{eq}) \times \hat{\mathbf{F}}_{m}$, $F_{sens} = p_s c_p C_D
c_{wave} /R \times \hat{F}_{sens}$.

The primitive and radiative equations in nondimensional form are
\begin{mathletters}
\begin{eqnarray*}
\left( \frac{D\hat{\mathbf{u}}}{D\hat{t}} \right) &=& -\frac{a^2}{L_{Ro}^2}
\left(\sin \theta \mathbf{k} \times \hat{\mathbf{u}} \right) - \frac{c_p}{R}~
\left(\hat{\nabla}\hat{\phi}\right) -\frac{C_D a}{H} \left(\frac{\partial \hat{\mathbf{F}}_{m} }{\partial \hat{p}}\right), \\
\left(\frac{\partial \hat{\phi}}{\partial \hat{p}}\right)&=& - \left(\frac{\hat{T}}{\hat{p}}\right), \\
\left(\hat{\nabla} \cdot \hat{\mathbf{u}}\right) & + & \left(\frac{\partial \hat{\omega}}{\partial \hat{p}}\right)= 0, \\
\left(\frac{D\hat{T}}{D\hat{t}}\right)&=& \frac{R}{c_p}
\left(\frac{\hat{T} \hat{\omega}}{\hat{p}}\right) + \frac{t_{wave}}{t_{rad}}
\left(\frac{\partial \hat{F}_{rad} }{\partial \hat{p}}\right) +
\frac{C_D a}{H} \left(\frac{\partial \hat{F}_{sens} }{\partial \hat{p}}\right) +
\frac{t_{wave}}{t_{rad}} \left(\frac{\partial \hat{F}_{conv}}{\partial \hat{p}}\right), \\
\left(\frac{\partial \hat{F}_{LW}^{\uparrow}}{\partial\hat{\tau}_{lw}}\right)&=& \hat{F}_{LW}^{\uparrow} - \hat{T}^4
\hspace{2cm} (0\leq \hat{\tau}_{lw}\leq\tau_{LW}), \\
\left(\frac{\partial \hat{F}_{LW}^{\downarrow}}{\partial \hat{\tau}_{lw}}\right)&=&-(\hat{F}_{LW}^{\downarrow}-
\hat{T}^4) \hspace{2cm} (0\leq \hat{\tau}_{lw}\leq\tau_{LW}), \\ 
\left(\frac{\partial \hat{F}_{SW}^{\downarrow}}{\partial
      \tau_{sw}}\right) & = &-\hat{F}_{SW}^{\downarrow} \hspace{2cm} (0\leq\hat{\tau}_{sw}\leq \gamma \tau_{LW}).
\end{eqnarray*}
\end{mathletters}

The surface energy budget in nondimensional form is
\begin{eqnarray*}
  \frac{t_{rad,s}}{t_{rad}} \left( \ddthat{\hat{T_s}}\right) & = &
  \hat{F}_{SW}^{\downarrow} |_{p_s} - (\hat{T}_s^4 -
  \hat{F}_{LW}^{\downarrow}|_{p_s} ) - \frac{t_{rad}}{t_{wave}}
  \frac{C_D a}{H} \hat{F}_{sens} |_{p_s},
\end{eqnarray*}
where $t_{rad,s} = \mathcal{C}/\sigma_{SB} T_{eq}^3$ is the surface
radiative timescale. Our assumption that $T_s$ is time-independent,
and $t_{rad,s}$ can be ignored, therefore breaks down when $t_{rad,s}
\gtrsim t_{rad}$.  In our reference simulations we assume a heat
capacity equivalent to that of a well-mixed water layer with depth
$3$ m. This means the surface thermal inertia, $\mathcal{C}$, is
actually large enough that $t_{rad,s} \sim t_{rad}$. To check if this
affects our results we recomputed the reference simulations in Figure
\ref{fig:three} with $\mathcal{C}$ reduced by a factor of 300, such
that $t_{rad,s} \ll t_{rad}$. We note that this heat capacity is far
less than realistic values for $\mathcal{C}$. At individual grid
points, the time-averaged surface temperature changes up to 2.5$\%$ in
the cool Earth-size scenario and up to $6\%$ for the hot
super-Earth. However, reducing $\mathcal{C}$ affects the phase curve amplitude of both runs by less
than $0.5\%$. This confirms that our choice of six nondimensional
parameters captures the dominant dynamics in the GCM simulations
(Fig.~\ref{fig:one}).

\section{Appendix: Tidally locked coordinate system}
\label{sec:appendix2}
A standard geographic coordinate system is defined via the radial
distance from a planet's center, $r$, the latitude, $\theta$, which is
the angle away from the equator, and the longitude, $\lambda$, which
is the angle about the planet's north pole. Atmospheres of fast
rotating planets are approximately symmetric around the axis of
rotation due to conservation of angular momentum, so their
time-averaged properties are often displayed as averages over
$\lambda$. Here we make use of the approximate symmetry of slowly
rotating tidally locked planets about the axis connecting the
substellar and antistellar points (Fig.~\ref{fig:zero}). We define the
tidally locked latitude, $\theta_{TL}$, as the angle away from the
terminator, and the tidally locked longitude, $\lambda_{TL}$, as the
angle about the substellar point. We choose $(\theta,\lambda) =(0,0)$
to coincide with the substellar point, and $(\theta_{TL},\lambda_{TL})
=(0,0)$ to coincide with the north pole (see top row in
Fig.~\ref{fig:zero}). For example, in tidally locked coordinates
$\lambda_{TL}=0$ and $90^\circ \geq \theta_{TL} \geq -90^\circ$
defines the arc that connects substellar and antistellar points via
the north pole.

To translate between standard and
tidally locked coordinates we first transform both spherical coordinate
systems into Cartesian coordinates, so that the north pole lies at
$(x,y,z)=(0,0,r)$ and the substellar point lies at $(x,y,z)=(r,0,0)$
\begin{eqnarray}
\label{eqn:B1}
x & = & r \cos\theta \cos\lambda, \\
y & = & r \cos\theta \sin\lambda,  \nonumber \\
z & = & r \sin\theta, \nonumber
\end{eqnarray}
and 
\begin{eqnarray}
\label{eqn:B2}
x & = & r \sin\theta_{TL}, \\
y & = & r \cos\theta_{TL} \sin\lambda_{TL}, \nonumber\\
z & = & r \cos\theta_{TL} \cos\lambda_{TL}. \nonumber
\end{eqnarray}
By combining equations \ref{eqn:B1} and \ref{eqn:B2} we can express
$\theta_{TL}$ and $\lambda_{TL}$ in terms of $\theta$ and $\lambda$:
\begin{eqnarray}
\label{eqn:B3}
\theta_{TL} & = & \sin^{-1}(\cos \theta \cos \lambda), \\
\lambda_{TL} & = & \tan^{-1}\left(\frac{\sin \lambda}{\tan \theta} \right). \nonumber
\end{eqnarray}
To plot GCM output in tidally locked coordinates we first
express the GCM output in terms of $\theta_{TL}$ and $\lambda_{TL}$,
and then linearly interpolate the output onto an evenly-spaced
$(\theta_{TL},\lambda_{TL})$ grid.

Transforming GCM wind velocities into tidally locked coordinates is
slightly more complicated. Horizontal winds are defined as $(u,v) \equiv \left( r \cos \theta
  \DDt{\lambda}, r \DDt{\theta} \right)$. We analogously define wind velocities
in a tidally locked coordinate system as
\begin{eqnarray*}
u_{TL} & \equiv & r \cos \theta_{TL} \DDt{\lambda_{TL}} \\
& = & r \cos \theta_{TL} \left( \partials{\lambda_{TL}}{\lambda}
  \DDt{\lambda}  +  \partials{\lambda_{TL}}{\theta} \DDt{\theta}\right) \\
& = &  \cos \theta_{TL} \left(
\partials{\lambda_{TL}}{\lambda} \frac{u}{\cos \theta} + \partials{\lambda_{TL}}{\theta} v \right),
\end{eqnarray*}
and
\begin{eqnarray*}
v_{TL} & \equiv & r \DDt{\theta_{TL}} \\
& = & r \left( \partials{\theta_{TL}}{\lambda} \DDt{\lambda} +\partials{\theta_{TL}}{\theta} \DDt{\theta} \right) \\
& = & \partials{\theta_{TL}}{\lambda} \frac{u}{\cos \theta} +\partials{\theta_{TL}}{\theta} v.
\end{eqnarray*}
We evaluate the partial derivatives
($\partial \lambda_{TL} / \partial \lambda, \partial
\lambda_{TL}/ \partial\theta,\partial \theta_{TL}/
\partial\lambda,\partial \theta_{TL} /\partial \theta$)
using equations \ref{eqn:B3}. The resulting
expressions are long and lead to little insight, so we omit them here.

\section{Appendix: Computing phase curves}
\label{sec:appendix3}
The area-averaged and observer-projected flux from a planet as seen by a distant observer is
\begin{eqnarray*}
  F(\xi) & = & \frac{\int_{-\pi/2}^{\pi/2}
  \int_{-\xi-\pi/2}^{-\xi+\pi/2} F_{LW}^{\uparrow}|_{p=0}
  \cos(\lambda+\xi) \cos^2(\theta) d\lambda d\theta}{\int_{-\pi/2}^{\pi/2}
  \int_{-\xi-\pi/2}^{-\xi+\pi/2} \cos(\lambda+\xi) \cos^2(\theta) d\lambda d\theta},
\end{eqnarray*}
where $\xi$ is the phase angle, i.e., the angle between the observer's
line-of-sight and the substellar point ($\xi=0$ at secondary eclipse,
$\xi=\pi$ at transit), and $F_{LW}^{\uparrow}|_{p=0}$ is the outgoing
thermal flux at the top-of-atmosphere. This equation is expressed in
standard latitude-longitude coordinates, and assumes that the orbit is
viewed edge-on \citep{cowan2008}. The planet's total flux
as seen by a distant observer is $\pi a^2 \times F(\xi)$.

The maximum thermal flux emitted by a planet corresponds to the flux
emitted by the dayside of a bare rock, which we call $F_{rock}$. We
compute $F_{rock}$ by setting the outgoing thermal flux equal to the
incoming stellar flux at every point, $F_{LW}^{\uparrow}|_{p=0} =
L_*(1-\alpha) \cos(\theta) \cos(\lambda)$, so $F_{rock} = 2/3\times
L_*(1-\alpha)$. We define the dayside-averaged
  observer-projected temperature of a bare rock as $\sigma_{SB}
  T_{rock}^4 = F_{rock}$. This temperature is related to the
  equilibrium temperature of a planet with effective heat transport
  via $T_{rock} = (8/3)^{1/4} T_{eq}$.

\section{Appendix: Finding all dimensional transformations that only
  affect one nondimensional parameter}
\label{sec:appendix4}

We illustrate how to find the set of all transformations on
the dimensional parameters that only vary one nondimensional
parameter. 
Our approach is similar to the commonly-used technique in
dimensional analysis of finding a complete set of
nondimensional parameters via matrix methods \citep{price2003}.
The method is general, but we illustrate it assuming
$(a,\Omega,T_{eq})$ are fixed to directly explain our parameter
choices in Table \ref{tab:params}b.
To transform the remaining dimensional parameters, we consider
multiplying each of them by a different constant,
\begin{align*}
(R',c_p',g',&p_s',\kappa_{SW}',\kappa_{LW}',C_D') = \\
& (C^{v_1} R,C^{v_2} c_p, C^{v_3} g, C^{v_4} p_s, C^{v_5}\kappa_{SW}, C^{v_6}\kappa_{LW}, C^{v_7} C_D),
\end{align*}
where a prime denotes a transformed parameter, $C$ is a constant, and
$(v_1,v_2,...)$ are different exponents. At the same time we 
want to keep all nondimensional parameters except one fixed. As an
example we consider all transformations that multiply $R/c_p$ by a
factor of $C$. This means
\begin{eqnarray*}
  \left( \frac{R}{c_{p}} \right)' = C \times \left( \frac{R}{c_{p}}
  \right) & \Rightarrow & C^{v_1} C^{-v_2} \left( \frac{R}{c_{p}}
  \right) = C \left( \frac{R}{c_{p}} \right)  \\
  & \Rightarrow & C^{v_1-v_2} = C^1 \\
  \left( \frac{a^2}{L_{Ro}^2} \right)' = C \times \left( \frac{a^2}{L_{Ro}^2} \right) &
  \Rightarrow & C^{-v_1+v_2/2} \left( \frac{a^2}{L_{Ro}^2}
  \right) = C^0 \left( \frac{a^2}{L_{Ro}^2} \right) \\
  & \Rightarrow & C^{-v_1+v_2/2} = C^0 \\
  \left( \frac{t_{wave}}{t_{rad}} \right)' = C \times \left(
    \frac{t_{wave}}{t_{rad}} \right)  & \Rightarrow &
  C^{-v_1-v_2/2+v_3-v_4} \left( \frac{t_{wave}}{t_{rad}}
  \right) = C^0 \left( \frac{t_{wave}}{t_{rad}} \right) \\
  & \Rightarrow & C^{-v_1-v_2/2+v_3-v_4} = C^0 \\
 & \vdots
\end{eqnarray*}
We consider the exponents of $C$ in the resulting equations. They form
a linear system of equations and can be written in matrix
form $A \mathbf{v} = (1,0,0,...)$ as
\begin{eqnarray*}
 \left(
\begin{array}{ccccccc}
  1 & -1 & 0 & 0 & 0 & 0 & 0 \\
  -1 & 1/2 & 0 & 0 & 0 & 0 & 0 \\
  -1 & -1/2 & 1 & -1 & 0 & 0 & 0 \\
  0 & 0 & 0 & 0 & 1 & -1 & 0 \\
  0 & 0 & -1 & 2 & 0 & 1 & 0 \\
  -1 & 0 & 1 & 0 & 0 & 0 & 1 \\
\end{array}
\right)
\left( \begin{array}{c}
v_1 \\ v_2 \\ v_3 \\ v_4 \\ v_5 \\ v_6 \\ v_7 
\end{array}\right)
& = &
\left( \begin{array}{c}
1\\0\\0\\0\\0\\0
\end{array}\right)
\end{eqnarray*}
The matrix columns correspond to the exponents of 
($R,c_p,g,p_s,\kappa_{SW},\kappa_{LW},C_D$), and the
matrix rows correspond to each nondimensional parameter. For example, the first
row of the matrix corresponds to $R/c_p$, and has only two non-zero entries
(corresponding to the exponents with which $R$ and $c_p$ appear in $R/c_p$).

This system of equations is underdetermined, that is, it has
infinitely many solutions (there are 6 rows/6 equations, but 7
columns/7 unknowns). All solutions can be expressed as a particular
solution $\mathbf{v}$ to the equation $A\mathbf{v}=(1,0,0,...)$, plus any
vector that lies in the kernel (nullspace) of $A$: $\{ \mathbf{v} + \mathbf{x}, ~\mathrm{where}~
\mathbf{x} ~\mathrm{satisfies}~ A\mathbf{x}=\mathbf{0} \}$. We express
the set of all solutions as
\begin{eqnarray*}
  \{\mathbf{v} + k \cdot \mathbf{x} \} & = & \left( -1, -2, -1-k, 1-k, -3+k, -3+k, k \right),
\end{eqnarray*}
where $k$ is an arbitrary number, and each entry of this vector
corresponds to the power to which $C$ is being raised for each
dimensional parameter. For example, to increase $R/c_p$ by
$C$, the first dimensional parameter, $R$, is multiplied by $C^{-1}$,
the second dimensional parameter, $c_p$, is multiplied by $C^{-2}$
etc.

To vary $R/c_p$ over the largest possible range compatible with the
dimensional constraints in Table \ref{tab:minmax} amounts to finding
the largest and smallest values of $C$ that are still consistent with Table
\ref{tab:minmax}, while $k$ is allowed to take on any value. We first
solve this problem by inspection, and then compare our parameter
choices to the values we get from numerical optimization. For our
reference case of a cool, Earth-sized planet we find
$k=0$ and $1/\sqrt{1.5} \lesssim C \leq 1$ (Table
\ref{tab:params}b).

We note that one can use this method similarly to find the set of all
transformations that leave all nondimensional parameters
invariant (see Table \ref{tab:params}a, Fig.~\ref{fig:one}). This set
of transformations is simply given by the kernel of $A$.


\bibliographystyle{apj}
\bibliography{./ZoteroLibrary_NoURLs}

%% the file, uncomment the \plotone, \plottwo, and
%% If you need a layout that cannot be achieved with \plotone or
%% \plottwo, you can invoke the graphicx package directly with the
%% \includegraphics command or use \plotfiddle. For more information,
%% please see the tutorial on "Using Electronic Art with AASTeX" in the
%% documentation section at the AASTeX Web site, http://aastex.aas.org/
%%
%% The examples below also include sample markup for submission of
%% supplemental electronic materials. As always, be sure to check
%% the instructions to authors for the journal you are submitting to
%% for specific submissions guidelines as they vary from

%% This example uses \plotone to include an EPS file scaled to
%% 80% of its natural size with \epsscale. Its caption
%% has been written to indicate that additional figure parts will be
%% available in the electronic journal.

\end{document}